\documentclass[2column]{aa}
\usepackage{natbib}
\usepackage{multirow}
\usepackage{color}
\usepackage[colorlinks=true]{hyperref}
\hypersetup{citecolor= blue, linkcolor=red, urlcolor=blue}

\usepackage[us,12hr]{datetime}
\usepackage{ulem}

\definecolor{deepmagenta}{rgb}{0.8, 0.0, 0.8}
\definecolor{carloscolor}{rgb}{0.2, 0.4, 0.8}

\begin{document}

   \title{Three-dimensional magnetic field structure of a flux emerging region in the solar atmosphere}
   \author{Rahul Yadav
          \inst{1}
          \and
          J. de la Cruz Rodr\'iguez
          \inst{1}
          \and
          C. J. D\'iaz Baso
          \inst{1}
          \and
          Avijeet Prasad
          \inst{2}
          \and
          \\
          Tine Libbrecht
          \inst{1}
          \and
          Carolina Robustini
          \inst{1}
          \and
          A. Asensio Ramos
          \inst{3}
          }

   \institute{Institute for Solar Physics, Dept. of Astronomy, Stockholm University, AlbaNova University Centre, SE-10691 Stockholm, Sweden
              \email{rahul.yadav@astro.su.se}
              \and Center for Space Plasma and Aeronomic Research, The University of Alabama in Huntsville, Huntsville, Alabama 35899, USA
              \and
              Instituto de Astrofisica de Canarias, Via Lactea s/n, E-38205 La Laguna, Tenerife, Spain
             }

   \date{Draft: compiled on \today\ at \currenttime~UT}
   
   \authorrunning{Yadav et al.}
   \titlerunning{}
 \abstract{We analyze high-resolution spectropolarimetric observations of a flux emerging region (FER) in order to understand its magnetic and kinematic structure. Our spectropolarimetric observations in the \ion{He}{I}~10830~\AA~spectral region of a FER are recorded with GRIS at the 1.5 m aperture GREGOR telescope. A Milne-Eddington based inversion code was employed to extract the photospheric information of the \ion{Si}{i} spectral line, whereas the \ion{He}{i} triplet line was analyzed with the Hazel inversion code, which takes into account the joint action of the Hanle and the Zeeman effect. The spectropolarimetric analysis of \ion{Si}{i} line displays a complex magnetic structure near the vicinity of FER, where a weak (350 – 600~G) and horizontal magnetic field was observed. In contrast to the photosphere, the analysis of the \ion{He}{i} triplet presents a smooth variation of the magnetic field vector (ranging from 100~G to 400~G) and velocities across the FER. Moreover, we find supersonic downflows of $\sim$40~km s$^{-1}$ appears near the foot points of loops connecting two pores of opposite polarity, whereas a strong upflows of 22~km\,s$^{-1}$ appears near the apex of the loops. At the location of supersonic downflows in the chromosphere, we observed downflows of 3~km\,s$^{-1}$ in the photosphere. Furthermore, non force-free field extrapolations were performed separately at two layers in order to understand the magnetic field topology of the FER. We determine, using extrapolations from the photosphere and the observed chromospheric magnetic field, that the average formation height of the \ion{He}{i} triplet line is $\sim$2~Mm from the solar surface. The reconstructed loops using photospheric extrapolations along an arch filament system have a maximum height of $\sim$10.5~Mm from the solar surface with a foot-points separation of $\sim$19~Mm, whereas the loops reconstructed using chromospheric extrapolations are around $\sim$8.4~Mm high from the solar surface with a foot-point separation of $\sim$16~Mm at the chromospheric height. The magnetic topology in the FER suggests the presence of small-scale loops beneath the large loops. Under suitable conditions, due to magnetic reconnection, these loops can trigger various heating events in the vicinity of the FER.}


%


   \maketitle
\section{Introduction}


The emergence of magnetic flux at the solar surface plays a vital role in understanding the formation of active regions, and in general the solar activity \citep{1985SoPh..100..397Z,vanDriel-Gesztelyi2015}. Magnetic flux emerging regions (FER) are also important to understand the dynamic coupling between different layers of solar atmosphere as they connect interior and outer solar atmosphere \citep{Cheung2014}. FERs, or in general active regions, are formed by the rise of flux tubes from the convection zone to the solar surface due to magnetic buoyancy or Parker instability \citep{1955ApJ...121..491P}. These flux tubes, when piercing the solar surface, form bipolar regions and generate a serpentine-like configuration over a wide range of spatial scales. On granular scales (1--2 Mm), due to granular convection, many small-scale bipolar or unipolar magnetic structures emerge or disappear \citep{2002SoPh..209..119B}. Whereas on large scales ($\sim$100 Mm), the interaction between photospheric motions and the cancellation or coalescense of small bipolar/unipolar region leads to the formation of active regions \citep{1995ApJ...441..886C, 2004LRSP....1....1F}.

In the chromosphere, emerged loops in a FER are commonly known as arch filament system (AFS). They appear as arch-like shape when observed using the line core filtergrams of H$\alpha$, Ca~\scriptsize{II}\normalsize{}~ H \& K and He~\scriptsize{I}\normalsize{}~10830 \AA. These emerged loops become dark due to their high density of chromospheric material, which leads to enhanced absorption under favorable conditions. Generally, the length of AFS or magnetic loops can lie between 20--30 Mm, whereas their width can reach a few mega-meters. The maximum height of AFS loop can varies between 5--25 Mm with a lifetime of about 30 minutes \citep{1967SoPh....2..451B}. Near both foot-points of AFS supersonic downflow (30--50 km s$^{-1}$), whereas near the middle part of loops strong upflows (2--20~km s$^{-1}$) were observed by various authors \citep{1967SoPh....2..451B, 2003Natur.425..692S,2018A&A...617A..55G}. For the structure and dynamics of AFS see a review by \cite{1993ASPC...46..471C} and references therein.

Photospheric magnetic field measurements, mainly dominated by the Zeeman effect, are commonly available from various space and ground based telescopes. However, similar measurements in the chromosphere and corona are challenging due to low magnetic field strength and the availability of few sensitive spectral lines \citep{2006RPPh...69..563S,2014A&ARv..22...78W}. Fortunately, the \ion{He}{i}~10830~\AA\ triplet line offers a unique diagnostics of the upper chromosphere. The  excitation of the \ion{He}{i} triplet line in the upper chromosphere is partially due to photoionisation from EUV radiation and to collisional excitation in regions with electron temperatures higher than 20,000 K \citep{1939ApJ....89..673G,1975ApJ...199L..63Z,1997ApJ...489..375A,2008ApJ...677..742C,2016A&A...594A.104L}. The magnetically sensitive \ion{He}{i}~10830 \AA~multiplet originates between the lower 2$^{3}$S$_{1}$ level and upper 2$^{3}$P$_{2, 1, 0}$ level. It consists of a blue component at 10829.0911~\AA, and two blended red components at 10830.2501~\AA\ and at 10830.3397~\AA.  
The linear polarization of the \ion{He}{i} triplet is dominated by atomic level polarization, and in the presence of a magnetic field this atomic level polarization is modified by the Hanle effect \citep{2002Natur.415..403T}. Circular polarization in the \ion{He}{i} 10830 \AA~lines is dominated by the longitudinal Zeeman effect, whereas the linear polarization is generally caused by the joint action of the transverse Zeeman effect and atomic level polarization even for the magnetic strength of $\sim$1~kG \citep{2007ApJ...655..642T}.
However, for stronger field strengths the linear polarization of \ion{He}{i} triplet is dominated by the transverse Zeeman effect. 

The helium triplet line has been utilized by distinct authors to investigate various chromospheric features. For example, magnetic loops \citep{2003Natur.425..692S,Xu2010}, supersonic downflows \citep{2007A&A...462.1147L}, the chromosphere above sunspots \citep{Schad2015,2016A&A...596A...8J}, flares \citep{2015ApJ...799L..25K,2015ApJ...814..100J}, spicules \citep{2005A&A...436..325L,2005ApJ...619L.191T,2010ApJ...708.1579C,2012ApJ...759...16M,2015ApJ...814..100J,2015ApJ...803L..18O}, prominences \citep{1981A&A...100..231B,2003ApJ...598L..67C,2006ApJ...642..554M,2014A&A...566A..46O,Gonz_lez_2015} and filaments \citep{2009A&A...501.1113K,2014A&A...561A..98S,2019A&A...625A.128D}. Furthermore, the \ion{He}{i} triplet line within the coronal loops can be utilized for coronal magnetic field measurements \citep{2016ApJ...833....5S,2018ApJ...865...31S}.

The first inference of magnetic field vector in FER in the upper chromosphere was presented by \cite{2003Natur.425..692S}, derived by inverting the full Stokes profiles of the \ion{He}{i}~10830 \AA~triplet. Assuming that the emerged loops are directed along the magnetic field lines, using the inferred magnetic field vector, they reconstructed full 3D structures of magnetic loops in the solar atmosphere. They reported that the reconstructed loops are $\approx$10~Mm high from the solar surface with strong downflows near the foot-points. Furthermore, the magnetic field strength decreased with height in both legs from 400 and 500~G at the chomospheric foot-points to below 50~G at the apex. Later, \cite{2009A&A...493.1121J} argued that they measured a magnetic field at a nearly constant height of $\approx$2.4~Mm, not along a magnetic loops. In order to check the arguments provided by \cite{2003Natur.425..692S} another study was performed in a different active region by \citet{Xu2010}. They also observed that the reconstructed loops are aligned with the long dark features visible in \ion{He}{I} triplet. To further remove the discrepancy about the height at which the \ion{He}{i} absorption take place, \cite{2011A&A...532A..63M} investigated the influence of the height on the strength of the Stokes Q and U profiles. They found that the linear polarization signals are well reproduced if the loop apex is higher than $\sim$5~Mm, and concluded that the dark features in FER are tracing the emerged loops. 

Although the investigations of FER at the photospheric layers are performed by various authors, spectropolarimetric analysis in the upper atmosphere are still limited (see \citealt{Cheung2014}, and references therein). Thus, in order to enhance our understanding regarding the FER in upper solar atmosphere we present the spectropolarimetric analysis of a FER. The main motivation of this article is to understand the kinematic and magnetic structure of a FER in the photosphere and the chromosphere. Moreover, to understand the magnetic topology in FER, the extrapolations are performed separately at two layers using the measured magnetograms. Even though the analysis of FER or active regions using a single layer (photosphere) extrapolations has been studied by various authors (\citealt{2012LRSP....9....5W}, and references therein), their investigation using two layers extrapolations is less explored. This analysis of FER using two layers extrapolation provide us several advantages. It can allow us to check the reliability of the extrapolations, to determine the relative formation height of spectral lines, and to test the consistency between the magnetic field observed at two different layers \citep{2012ApJ...748...23Y}.


This article is organized in the following manner. The overview of observations are described in Sect.~\ref{sec-observation}.  The analysis of spectropolarimetric data is discussed in Sect.~\ref{sec_inversion}. The extrapolations using the observed magnetograms are described in Sect.~\ref{sec_nfff}. Finally, the results and discussions are presented in Sections~\ref{sec_results} and \ref{sec_conclusion}, respectively.

\section{Observations}
\label{sec-observation}
\begin{figure}[!t]
\centering
\includegraphics[width=0.5\textwidth]{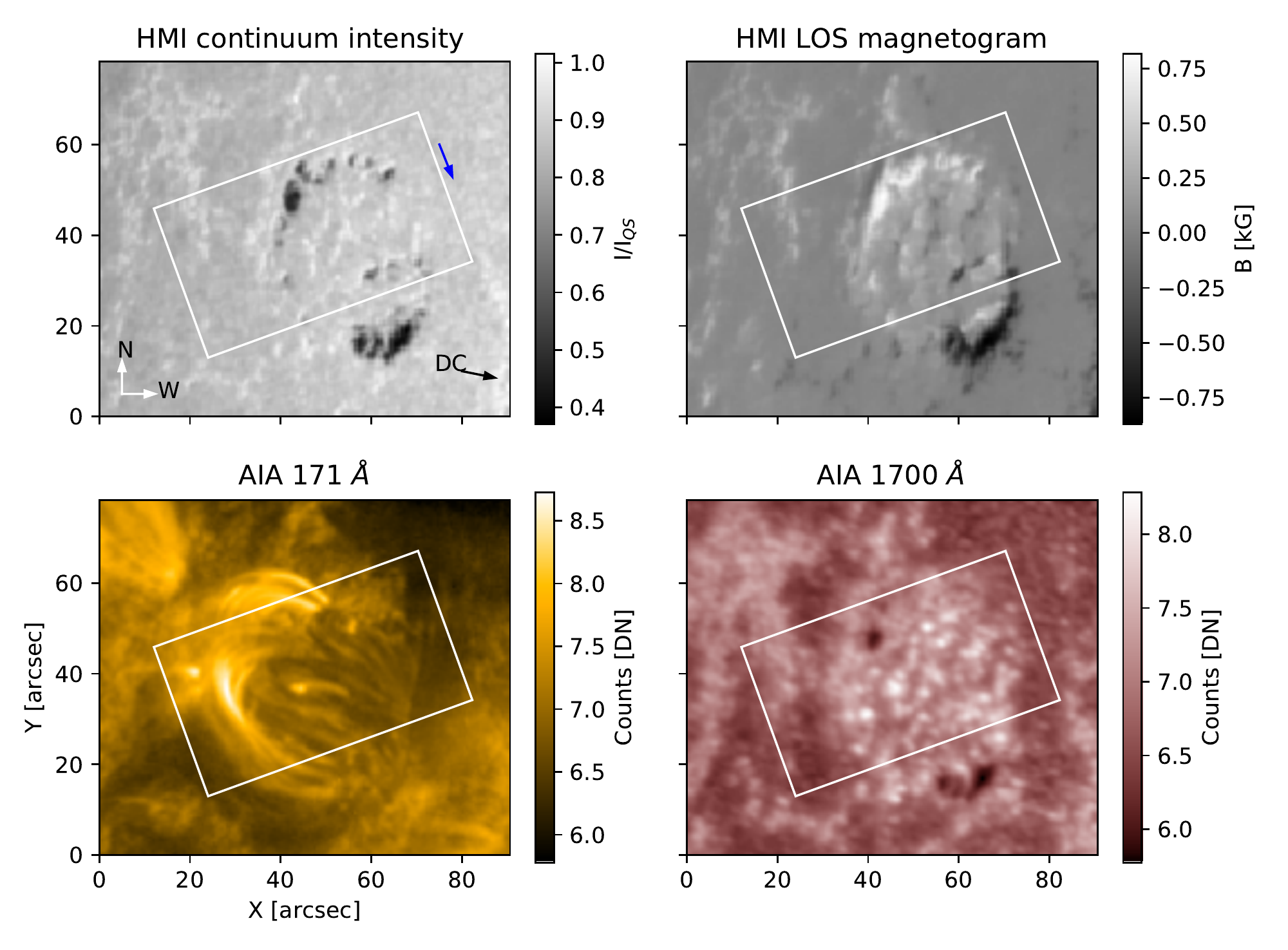}
\caption{Overview of the observations at 10:36 UT on 3rd June, 2015. Top panels: HMI continuum image and LOS magnetogram of FER. Bottom panels: AIA 171~\AA$ $ and 1700~\AA$ $ images of FER. A white box rectangle outlines the FOV covered by GRIS. A black and blue arrow in the top left panel indicate the direction of solar disc center and the scanning direction of the slit, respectively. Solar north and west directions are indicated by N and W. }
\label{overview}
\end{figure}

A flux emerging region, located near the solar limb, X=$-$789\arcsec \& Y=224\arcsec ($\mu$=0.52), was recorded between 10:36 and 10:48~UT on 3rd June, 2015. The observations were performed using the GREGOR Infrared Spectrograph (GRIS, \citealt{2012AN....333..872C}) at the 1.5 m aperture GREGOR telescope at Observatorio del Teide, Tenerife \citep{2012AN....333..796S}. The GRIS recorded the full Stokes parameters ($\mathrm{I}$, $\mathrm{Q}$, $\mathrm{U}$, and $\mathrm{V}$) along the spectrograph slit. It covers the spectral region spanned from 10822~\AA\ to 10836~\AA$ $, which contain some photospheric lines (Si~\scriptsize{I}\normalsize{} and Ca~\scriptsize{I}\normalsize{}) and the chromospheric He~\scriptsize{I}\normalsize{} 10830~\AA$ $ triplet line. The spectral sampling along the wavelength axis was 18~m\AA$ $ per pixel. The spatial pixel size and step size along the slit was 0\arcsec.135. The spatial scan was performed along the perpendicular direction of the slit in total 250 steps, resulting a field-of-view (FOV) of 34\arcsec $\times$ 60.5\arcsec. The observed FOV is outlined in Fig.~\ref{overview}.

\subsection{Data reduction}
Standard calibration routines were applied to the raw data, which include the correction for dark current, flat-field, polarimetric calibration, wavelength calibration and correction of residual fringes \citep{2003SPIE.4843...55C}. In order to improve the signal-to-noise ratio (SNR), we averaged 2$\times2$ spatial pixels and 2 spectral pixels, which results in the spatial and spectral resolution of 0\arcsec.27 and 36 m\AA$ $ per pixel, respectively. After these corrections, we employed the PCA de-noising technique \citep{2008A&A...486..637M} to further improve the SNR in Stokes parameters, which yields the noise level in Stokes ($\mathrm{Q}$, $\mathrm{U}$, and $\mathrm{V}$) of the order of (4, 4, 6)$\times10^{-4}$I$_{c}$, where I$_{c}$ is the continuum intensity. A relative velocity calibration was performed by fitting an averaged Stokes $\mathrm{I}$ profiles of Si line, which was obtained by taking average of intensity profiles over a quiet-sun region in the scanned FOV.

Although the above procedure reduces the noise in the observed data significantly, we noticed that there were some spurious short-period and small amplitude fringes still present over the entire spectral range, mainly in the Stokes~$\mathrm{I}$ profiles. One of the reasons for the appearance of these fringes could be due to multiple reflections on the surface of the optical devices placed in the optical path. Thus, in order to remove the contribution of these fringes in Stokes~$\mathrm{I}$ profiles we employed the Relevent Vector Machine method (RVM; \citealp{Tipping00therelevance}). This method, described in detail in \cite{2019A&A...625A.128D}, works in the following manner. The original signal, in our case the intensity spectrum, is decomposed into two components: one without fringes and another only with fringes. These two components are represented by two dictionaries (set of functions) of different properties. The first dictionary is made of Gaussian functions with different widths centered at several wavelengths positions across the spectrum. On the other hand, the second dictionary is made of sine and cosine functions of different periods. Then, the observed spectrum is fitted by taking the coefficients of these two dictionaries as free parameters (see Equation 4 in \citealt{2019A&A...625A.128D}). At last, the first dictionary made of Gaussian functions gives the spectrum free from fringes, whereas the second dictionary shows the presence of fringes in the spectrum.

\label{fig1}

\begin{figure}[!h]
\centering
\includegraphics[width=0.4\textwidth]{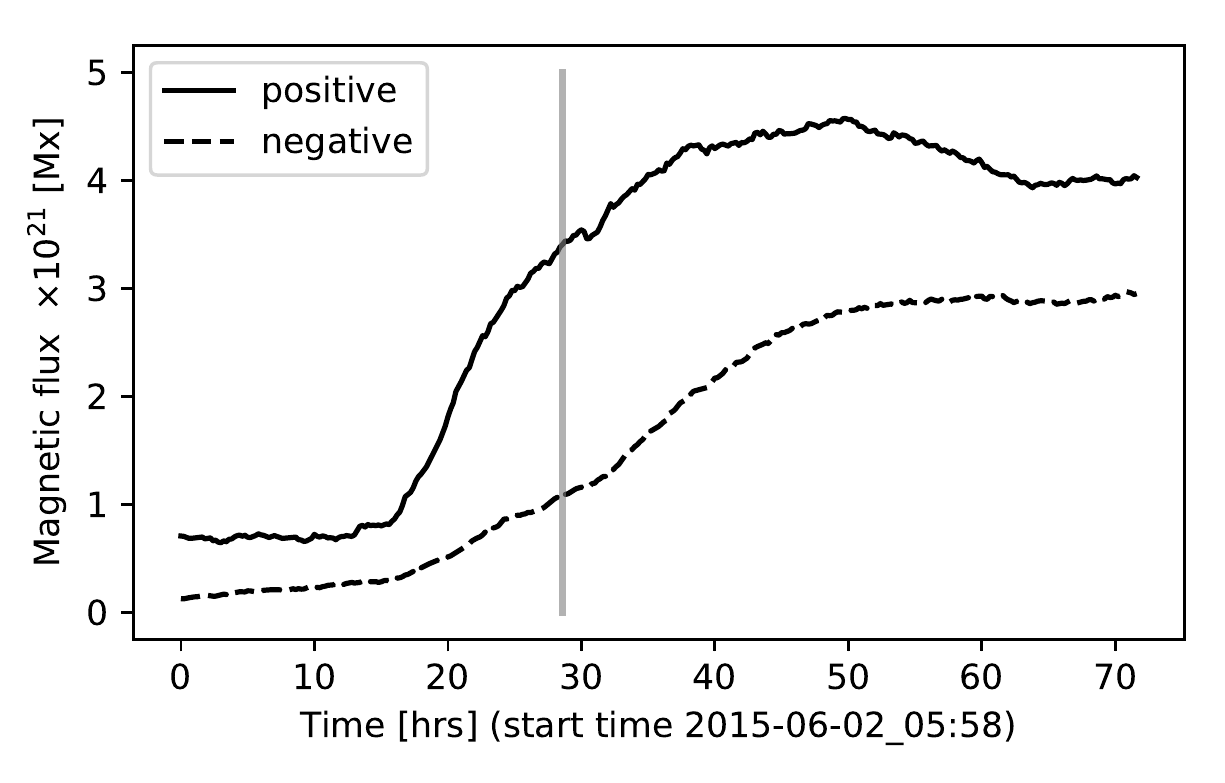}
\caption{The temporal evolution of positive and negative fluxes in FER is shown by solid and dashed line, respectively. A vertical gray solid line indicates the time of observations with GREGOR.}
\label{flux}
\end{figure}

\begin{figure}[!h]
\centering
\includegraphics[width=0.8\linewidth]{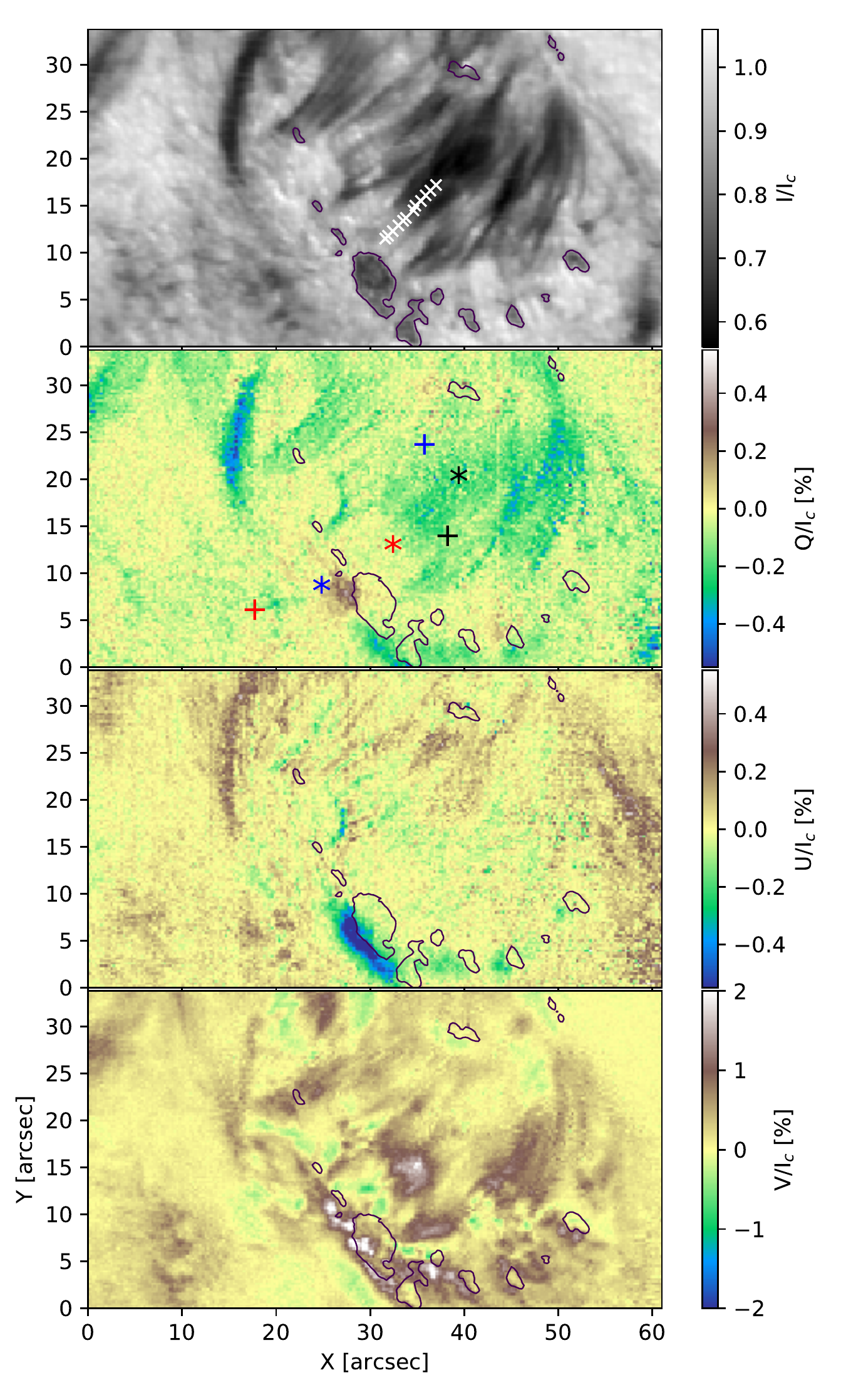}
\caption{Stokes $\mathrm{I}$, $\mathrm{Q}$, $\mathrm{U}$, and $\mathrm{V}$ maps (top to bottom) of the FOV observed with GRIS/GREGOR. Closed contours represents the pores shown in Fig.~\ref{overview}. Three plus and asterisk symbols in a panel refer to the inverted pixels shown in Fig.~\ref{1c_bestfit} and \ref{2c_bestfit}, respectively. In the top panel, white cross symbols indicate the positions of the Stokes I profiles shown in Fig.~\ref{stokesi_profiles}.}
\label{signal_iquv}
\end{figure}
\subsection{Overview of observations}
Figure~\ref{overview} shows an overview of the observation. The top and bottom panels show contextual images obtained with the Helioseismic and Magnetic Imager (HMI, \citealt{2012SoPh..275..207S}) and the Atmospheric Imaging Assembly (AIA, \citealt{2012SoPh..275...17L}), both are on board the Solar Dynamic Observatory (SDO, \citealt{2012SoPh..275....3P}). A white rectangle superimposed on the images represents the observed FOV. The HMI and AIA maps clearly show the complex structure of the FER. The sequential images obtained from HMI line-of-sight (LOS) magnetogram reveal patches of opposite polarities, moving, merging  and evolving with time, which leads to the enhancement in magnetic field strength and the development of pores. In AIA EUV 171~\AA$ $ map, small-scale loop connecting the emerging bipoles can be clearly identified. During flux emergence, various brightening events are noticed in different SDO/AIA filtergrams (304, 1600 and 1700~\AA$ $). For instance, as shown in Fig.~\ref{overview}, AIA 1700~\AA~filtergram reveal several transient brightening events mainly located near the location of mixed polarities.

In order to study the history of the FER, we estimated the magnetic flux rate of the FOV by integrating the positive and negative polarity fluxes separately. We have analyzed their temporal evolution a few days before and after the observations performed with GRIS. For this purpose, we have adopted the magnetic parameters obtained from the Space-weather HMI Active Region Patch (SHARP,  \citealt{2014SoPh..289.3549B}), which is derived from full-disc HMI data. The SHARPs images are deprojected to the heliographic coordinates with a Lambert (cylindrical equal area; CEA) projection method. The calculated magnetic flux rate for positive (black solid line) and negative (dashed black line) polarities, using the B$_z$ component of the magnetic field strength, are depicted in Fig.~\ref{flux}. We summed only those pixels with $|$B$_z|$ values larger than 30~G neglecting the contribution of weak signals \citep{2014SoPh..289.3483H}. During the flux emergence period, the flux of either polarity has increased up to $\sim$10$^{21}$~Mx. Furthermore, for either polarity, the estimated flux emergence rate during the GRIS observation turn out to be $\sim$10$^{16}$~Mx~s$^{-1}$. It is evident from Fig.~\ref{flux} that the spectropolarimetric observations with GRIS/GREGOR were performed roughly in the middle part of the flux emerging period, which is represented by a solid vertical gray line. 

Figure~\ref{signal_iquv} shows the four Stokes maps observed across the \ion{He}{i} line. These maps were generated near the core of the red component of \ion{He}{i} triplet. All maps were normalized with the continuum value of the quiet-sun region. The dark elongated features connecting pores of opposite polarities are illustrated in the top panel of Fig.~\ref{signal_iquv}. These loops are usually seen above the FER in chromospheric lines because the dense and cool plasma along the loop absorb the radiation coming from below. It is also evident from  Fig.~\ref{signal_iquv} that Stokes $\mathrm{Q}$ and $\mathrm{V}$ have sufficient signal in the loops, and thus they can be utilized to infer the magnetic field vector. Our analysis of the spectropolarimetric data is described in the following section.

\begin{figure*}[!h]
\centering
\includegraphics[width=0.95\textwidth]{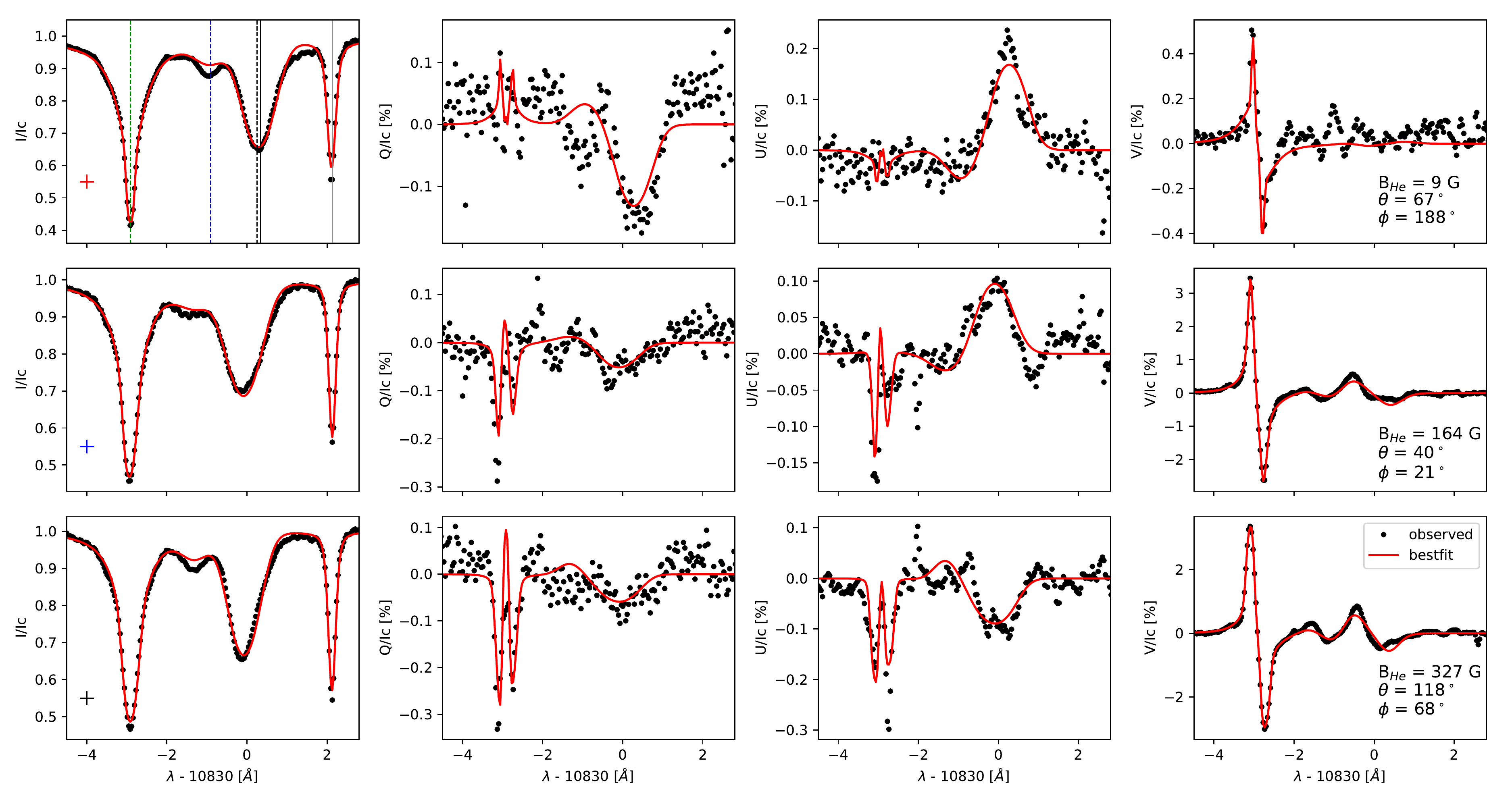}
\caption{The dark dotted and red solid lines represent the observed and best fitted Stokes profiles, respectively. All observed profiles were fitted with a single slab of model atmosphere and the inferred magnetic parameters (using He triplet) are given in the right column. The location of the observed Stokes profiles, in each panel, is indicated by plus signs of different colors in Fig. \ref{signal_iquv}. In the top left panel (Stokes I profile), the dashed green and solid gray line represent the \ion{Si}{i}~10827~\AA~and the telluric line (10832~\AA). The rest wavelength of blue (dashed blue) and two blended red components (black dashed and solid) of He I triplet lines are indicated by vertical lines.}
\label{1c_bestfit}
\end{figure*}

\begin{figure*}[!h]
\centering
\includegraphics[width=0.95\textwidth]{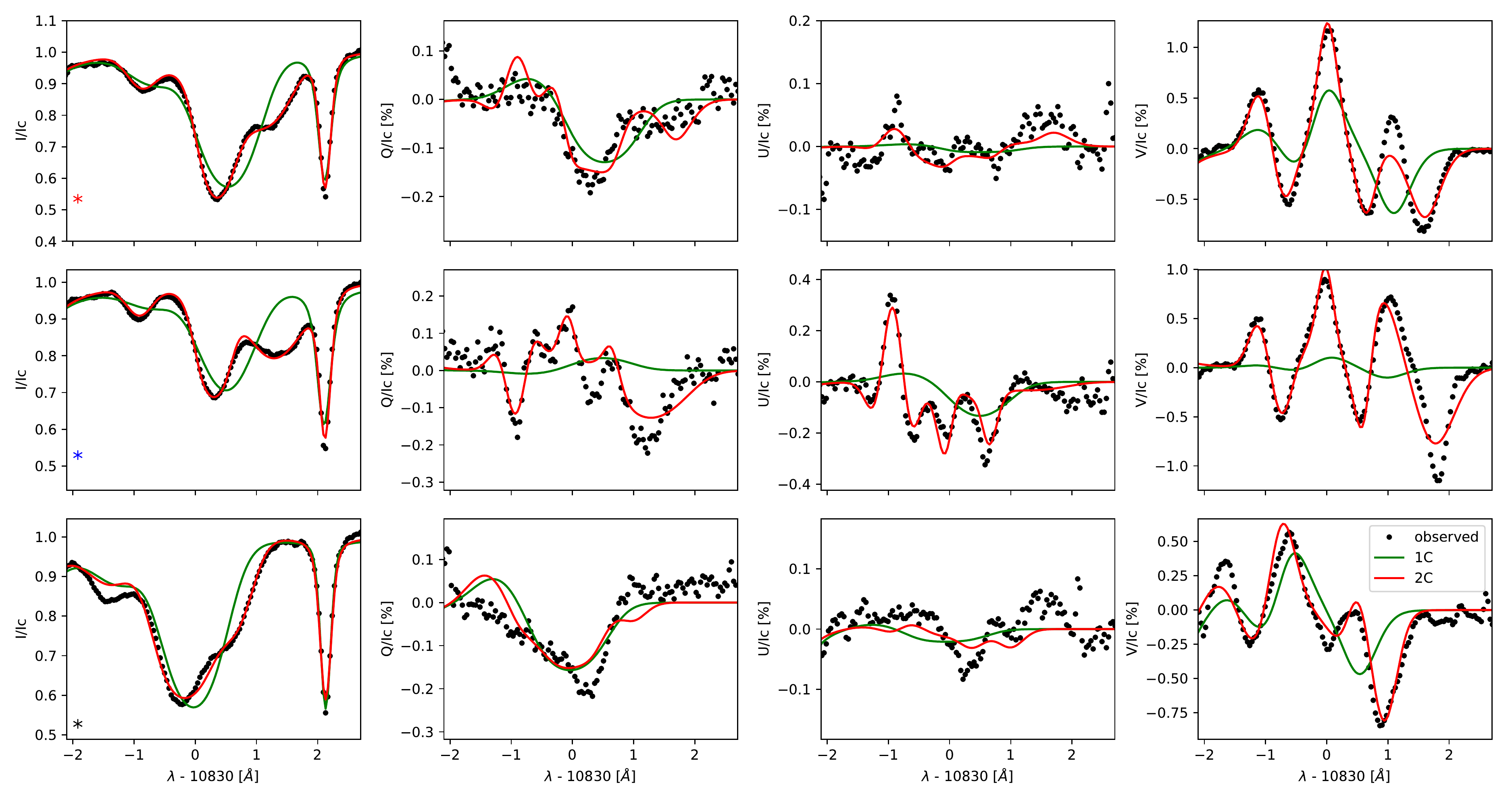}
\caption{Same as Fig \ref{1c_bestfit}, but the observed Stokes profiles were fitted using a single slab (solid green line; 1C) and double slab (solid red line; 2C) model atmosphere. The location of the observed Stokes parameters are indicated by 'asterisk' signs in Fig. \ref{signal_iquv} and the retrieved magnetic field vectors and LOS velocities are listed in Table \ref{tab-2slab}}
\label{2c_bestfit}
\end{figure*}

\begin{figure}[!h]
\centering
\includegraphics[width=0.45\textwidth]{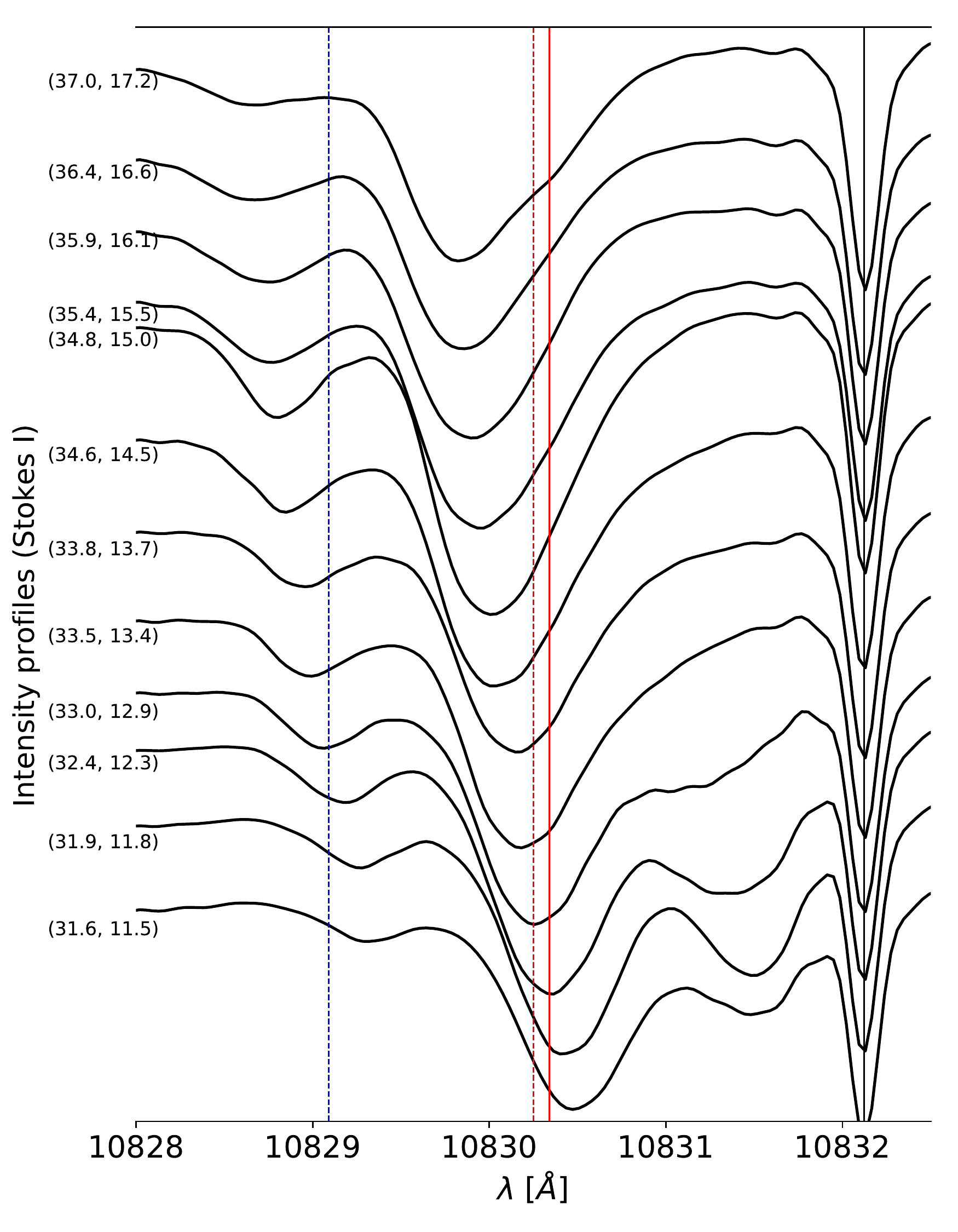}
\caption{Stokes I profiles along a loop. At the foot-point (lower profile) two velocity components can be noticed. The blue dashed, red dashed, solid red and solid black lines represent the rest wavelength of the \ion{He}{i}~triplet lines and the telluric line, respectively. The numbers in parentheses denote the position (in arcsec) of the pixels, which is indicated by white cross symbols in Fig. \ref{signal_iquv}.}
\label{stokesi_profiles}
\centering
\end{figure}

\begin{table*}[!t]
\centering
\caption{Parameters retrieved by a single slab and double slab model atmosphere. The location of inverted pixels are indicated by asterisk symbols in Fig \ref{signal_iquv}, whereas the observed and fitted profiles are shown in Fig. \ref{2c_bestfit}.}
\label{tab-2slab}
\resizebox{0.9\textwidth}{!}{%
\begin{tabular}{ccccc|cccc|cccc}
\hline

 & \multicolumn{4}{c}{1 component} & \multicolumn{8}{c}{2 component} \\

 & \multicolumn{1}{c}{} &  &  &  & \multicolumn{4}{c}{Slab1} & \multicolumn{4}{c}{Slab2} \\
 & B [G]& $\theta~[^{\circ}]$ & $\phi~[^{\circ}]$ & V$_{los}$ [{km~s}$^{-1}$]& B [G] & $\theta~[^{\circ}$]& $\phi~[^{\circ}$] & V$_{los}$ [{km~s}$^{-1}$]& B [G] & $\theta~[^{\circ}$] & $\phi~[^{\circ}$] & V$_{los}$ [{km~s}$^{-1}$] \\
 \hline 
\color{red}{*} & 307 & 68.1 & 94 & 6.2 & 829 & 68.9 & 32.3 & 20 & 402 & 57.7 & 45.9 & -0.9 \\
\color{blue}{*}  & 483 & 53.9 & 19 & 6 & 654 & 44.3 & 48.1 & 27.1 & 760 & 75.1 & 95.6 & -0.1 \\
\color{black}{*}  & 241 & 117 & 42.4 & -9.1 & 343 & 60 & 42.4 & -12.8 & 515 & 56.7 & 58.3 & 10 \\
\hline

\end{tabular}%
}
\end{table*}
\section{Analysis of Spectropolarimetric data}
\label{sec_inversion}
We analyze the spectropolarimetric data of \ion{Si}{i} 10827~\AA$ $ and \ion{He}{i} 10830~\AA$ $ lines in order to infer the physical properties of FER in the photosphere and chromosphere, respectively. 

\subsection{Inversion of \ion{Si}{I} 10827~\AA~line}
We first inferred the thermodynamic and magnetic parameters of photosphere using the \ion{Si}{i} 10827 \AA~line. All the Stokes profiles are inverted using the SPIN code \citep{2017SoPh..292..105Y}, which is based on the analytic solution of the radiative transfer equation for polarized radiation in a Milne-Eddington (M-E) model atmosphere \citep{1977SoPh...55...47A,2003isp..book.....D}. The code fits the Stokes profiles by varying the eight free parameters of a model atmosphere: the magnetic field strength (B), its inclination ($\theta$) and azimuth angle ($\phi$), LOS velocity (V$_{los}$), the line-to-continuum absorption ratio ($\eta_0$), the Doppler width ($\Delta\lambda_D$), damping constant ($a$), source function (S$_0$) and its gradient (S$_1$). Before the inversion, all four Stokes profiles were normalized with the averaged quiet-sun continuum intensity. Then, from the \ion{Si}{i} line inversion, we retrieve the magnetic field vector and LOS velocity in the photosphere. Generally, the inferred transverse components of magnetic field remain unchanged under a rotation of 180$^{\circ}$. This is also known as the 180$^{\circ}$ azimuthal ambiguity of the Zeeman effect \citep{1897ApJ.....5..332Z}.
Thus, in order to get the correct orientation of the magnetic field vector, this ambiguity has to be resolved before further investigation. For this purpose, different methods are available, some of them are described in \citet{Metcalf2006}. We adopted the automated ambiguity resolution code\footnote{Available at https://www.cora.nwra.com/AMBIG/} \citep{2014ascl.soft04007L}, which is based on minimum energy method \citep{1994SoPh..155..235M}. After this 180-ambiguity correction, we transform the magnetic field vector inferred in LOS frame to the solar local reference frame using the transformation matrix given by \cite{1990SoPh..126...21G}. 

\subsection{Inversion of \ion{He}{i} 10830~\AA~line}
\label{sec_helium_inversion}
The He triplet lines were inverted using the new version of Hazel inversion code\footnote{Available at https://github.com/aasensio/hazel2} \citep{2008ApJ...683..542A}. Unlike the previous version, the latest one employed here can synthesize and invert multiple spectral lines simultaneously. It can synthesize photospheric lines under the assumption of local thermodynamic equilibrium, chromospheric lines (for example, \ion{He}{i} triplet) under the multi-term approximation and a few telluric lines using the Voigt function. The code also uses the emergent intensity from the photosphere, derived using the synthetic module of the SIR inversion code \citep{1992ApJ...398..375R}, as an input to the chromospheric model atmosphere. The approach of coupling the photospheric radiation to the chromosphere can generate more realistic model atmosphere compared to the previous version of the Hazel code. 

Even though the stratified model parameters of the photosphere can be retrieved from the Hazel code, we employed a M-E based inversion code to invert the \ion{Si}{i} line because it provides us with averaged and height independent model parameters, which are relatively easy to interpret. 

To synthesize the Stokes profiles across \ion{He}{i} 10830 triplet lines, the Hazel code considers a constant-property slab model of plasma, located at a height '$h$' above the visible solar surface. It considers that all atoms inside this slab of optical thickness $\tau$ are illuminated from below by the photospheric solar continuum radiation field and the slab is placed in the presence of a deterministic magnetic field strength. The parameters that describe the slab model are the optical depth ($\tau$) at the core of the red component, the damping parameter ($a$), the Doppler width ($\Delta \lambda_D$), the LOS velocity of the plasma ($V_{LOS}$), the height of the slab from the solar surface, the heliocentric angle, and the three components of the magnetic field vector. Generally, the anisotropic radiation pumping produces population imbalances and quantum coherences between pairs of magnetic sublevels of the \ion{He}{i} atoms \citep{2007ApJ...655..642T}. In addition to this, these levels get influenced by the presence of magnetic field due to the Zeeman and the Hanle effect \citep{2002Natur.415..403T}. As a consequence of atomic polarization, Zeeman and Hanle effect, the emergent radiation is polarized. To calculate the emergent Stokes profiles ($\mathbf{I}$ = {I, Q, U, V}), the code uses the analytical solution of the radiative equations for a slab of constant properties \citep{2008ApJ...683..542A}, 
\begin{equation}
\label{slab-eq}
    \mathbf{I}(\tau) = \mathbf{I_{0}} \cdot e^{-\mathbf{K^{\ast}} \cdot \tau} + [\mathbf{K^{*}}]^{-1}(1-e^{-\mathbf{K^{*}}\cdot\tau})\mathbf{S},
\end{equation}
where $\mathbf{I_{0}}$ is the Stokes vector that illuminates the slab’s boundary from below, $\mathbf{K^{*} = K}$/$\eta_I$ and $\mathbf{S} = \boldsymbol{\epsilon}$/$\eta_I$. $\mathbf{K}$ and $\mathbf{S}$ are the propagation matrix and the source function, respectively. $\boldsymbol{\epsilon}$ and $\eta_I$ represent coefficients of the emission vector and  the coefficient of the propagation matrix, respectively.
For more details regarding the quantum theory of spectral line polarization and the solution of above equation see the monograph by \citet{2004ASSL..307.....L}. 

The Stokes I profiles in 10830~\AA{}, shown in Fig.~\ref{1c_bestfit}, illustrate that the \ion{Si}{i} 10827~\AA$ $ line blends with the blue component of the \ion{He}{i} triplet, whereas the red component blends with a telluric line (10832.3~\AA{}). Thus, in order to achieve a better fitting of profiles and to infer reliable model parameters, along with the \ion{He}{i} triplet lines, we also take into account the \ion{Si}{i} and the blended telluric line.
The following describes the configuration of Hazel code and the inversion scheme employed to invert the observed Stokes profiles. All pixels are set to invert using three inversion cycles. The Harvard-Smithsonian Reference model Atmosphere (HSRA; \citealt{1971SoPh...18..347G}) model is used as an initial model atmosphere for the \ion{Si}{i} spectral line. For all the inversion cycles we use 3 nodes in temperature, 2 nodes in all the components of the magnetic field vector, 2 nodes in line-of-sight velocity and 1 node in microturbulence. For the \ion{Si}{i} line we only considered one component of model atmosphere. Whereas, for the \ion{He}{i} triplet line, as a first exercise, all the pixels are inverted using a single slab model atmosphere placed at a fixed height of 3\arcsec~from the solar surface. In addition to this, we assume that the telluric line is well modelled by the Voigt function. The above configuration can  invert the whole spectral region around 10830~\AA\ and can be summarized as the light emerged from the photosphere is passing to the chromosphere and finally, absorbed by the telluric contamination.

In order to reduce the number of free parameters, we fixed only the damping parameter ($a$) to 10$^{-4}$ value. For the He triplet lines, \citet{2004A&A...414.1109L} have also demonstrated that the inferred parameters are not much affected by fixing $a$. In the first cycle, only thermodynamic parameters ($\Delta \lambda_D$, $\tau$, and V$_{LOS}$) are allowed to fit the intensity profile by fixing B$_x$, B$_y$, and B$_z$. In our case, generally, a single inversion cycle is sufficient to retrieve the thermodynamic parameters. In the next two cycles, where the outputs of the previous cycle are considered as the initial model input, we also allow the magnetic parameters to fit the polarized signals. Further, to avoid the possibility of local minimum of the $\chi^2$, we performed the inversion several times using different Stokes weights and initial guess models. As an example, Fig.~\ref{1c_bestfit} shows the observed Stokes profiles and the best fitted ones, retrieved using the Hazel code. 
Figures~\ref{1c_bestfit} and \ref{2c_bestfit} illustrate how the shape of the Stokes $\mathrm{Q}$ and $\mathrm{U}$ profiles across \ion{He}{i} triplet lines are generated due to both the Zeeman and the Hanle effect. A single lobe in Stokes $\mathrm{Q}$ and $\mathrm{U}$ indicates the presence of atomic polarisation. The amplitude of this lobe decreases in the presence of magnetic field due to the Hanle effect, which can be noticed in Fig.~\ref{1c_bestfit}. In strong field regions, three lobes signature of the Zeeman effect in Stokes $\mathrm{Q}$ and $\mathrm{U}$ can also be noticed in the middle panel of Fig.~\ref{2c_bestfit}.

Although most of the pixels are well fitted by a single slab model, we noticed that for some pixels the one slab model is not suitable to reproduce the observed Stokes profiles of the \ion{He}{i} triplet line. The signature of two-components is challenging to identify in Stokes $\mathrm{Q}$, $\mathrm{U}$ and $\mathrm{V}$ due to low signal-to-noise ratio and their complex shapes. Nevertheless, they can be identified in the Stokes I profiles, as shown in Fig.~\ref{2c_bestfit} \&~\ref{stokesi_profiles}. Generally, the two components have fast and slow velocities. As an example, Fig.~\ref{stokesi_profiles} demonstrates the appearance of single and double velocity components in the Stokes I profile, one of them (the slow component) is located close to the rest wavelength. We identify these pixels using a threshold value of $\chi^2$, obtained after fitting the Stokes I profiles using one and two components of model atmosphere. They are mainly located near the foot-points of loops or near the location of mixed bipolar fields. 
For these pixels, we take into account the two slab model atmosphere. 

\begin{figure*}[!h]
\centering
\includegraphics[width=0.9\textwidth]{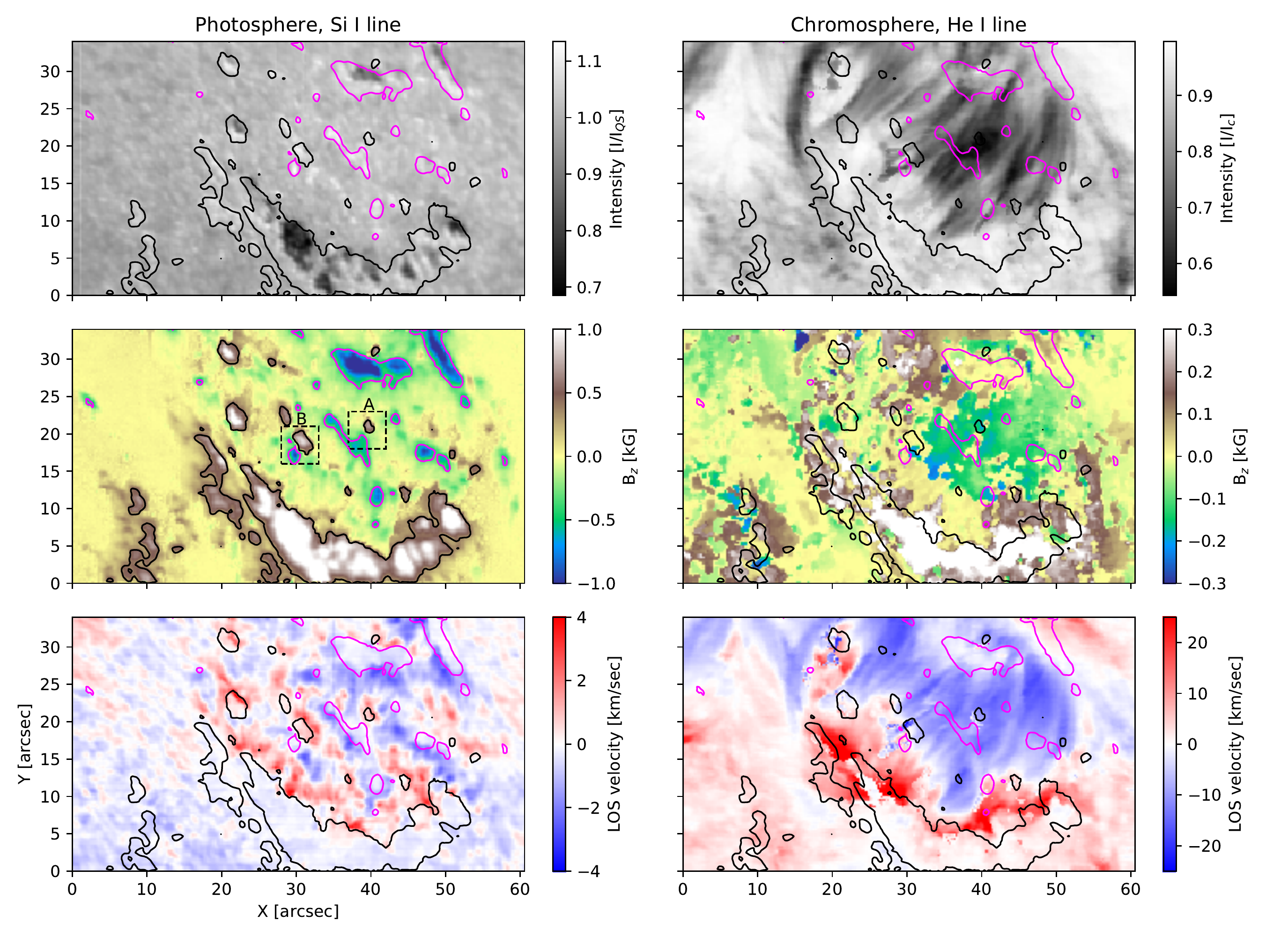}
\caption{Left panel: inferred photospheric parameters using M-E inversion of Si~\scriptsize{I}\normalsize{}~line. Right panel: inferred parameters obtained by inverting He triplet lines using Hazel code. The magenta and black closed contours in each map represent the negative and positive polarity of the magnetic field observed in the photosphere, respectively.  Square boxes (A and B) indicate the location of mixed polarity region.}
\label{photo_chromo}
\end{figure*}

In the Hazel code, two model atmospheres can be combined with the filling factor or as a stacked atmospheres. In the latter case two slab models are placed on top of each other, whereas in the former case they are combined with a filling factor in a single pixel. It has been demonstrated that in a FER various magnetic loops carrying helium plasma can appear at different heights \citep{2003Natur.425..692S,Xu2010}. Under this scenario, the stacking approach can be more realistic. Thus, in order to fit the two-components profiles we considered stacking approach instead of the filling factor. We placed the first (slab1) and the second slab (slab2) at the height of 2\arcsec~and 3\arcsec~from the solar surface, respectively. In this configuration, the light coming from photosphere first pass through slab1, then slab2 and finally is absorbed by the Earth's atmosphere (see also Eq.~4 in \citealt{2017A&A...598A..33L}). As an example, the observed and best fitted profiles are shown in Fig.~\ref{2c_bestfit}. It also shows that a single slab model is unable to fit the observed Stokes profiles, however, using two slab model a reasonably good fits can be achieved. For the selected pixels, the inferred parameters using a single and double slab models are listed in Table~\ref{tab-2slab}. The retrieved atmospheric parameters of FER in the photosphere and chromosphere are shown in Fig. \ref{photo_chromo}. After the inversion of full FOV, the next step is to remove the ambiguity in the retrieved azimuthal angle, which is discussed in the following section.

\begin{figure}[]
\centering
\includegraphics[width=0.49\textwidth]{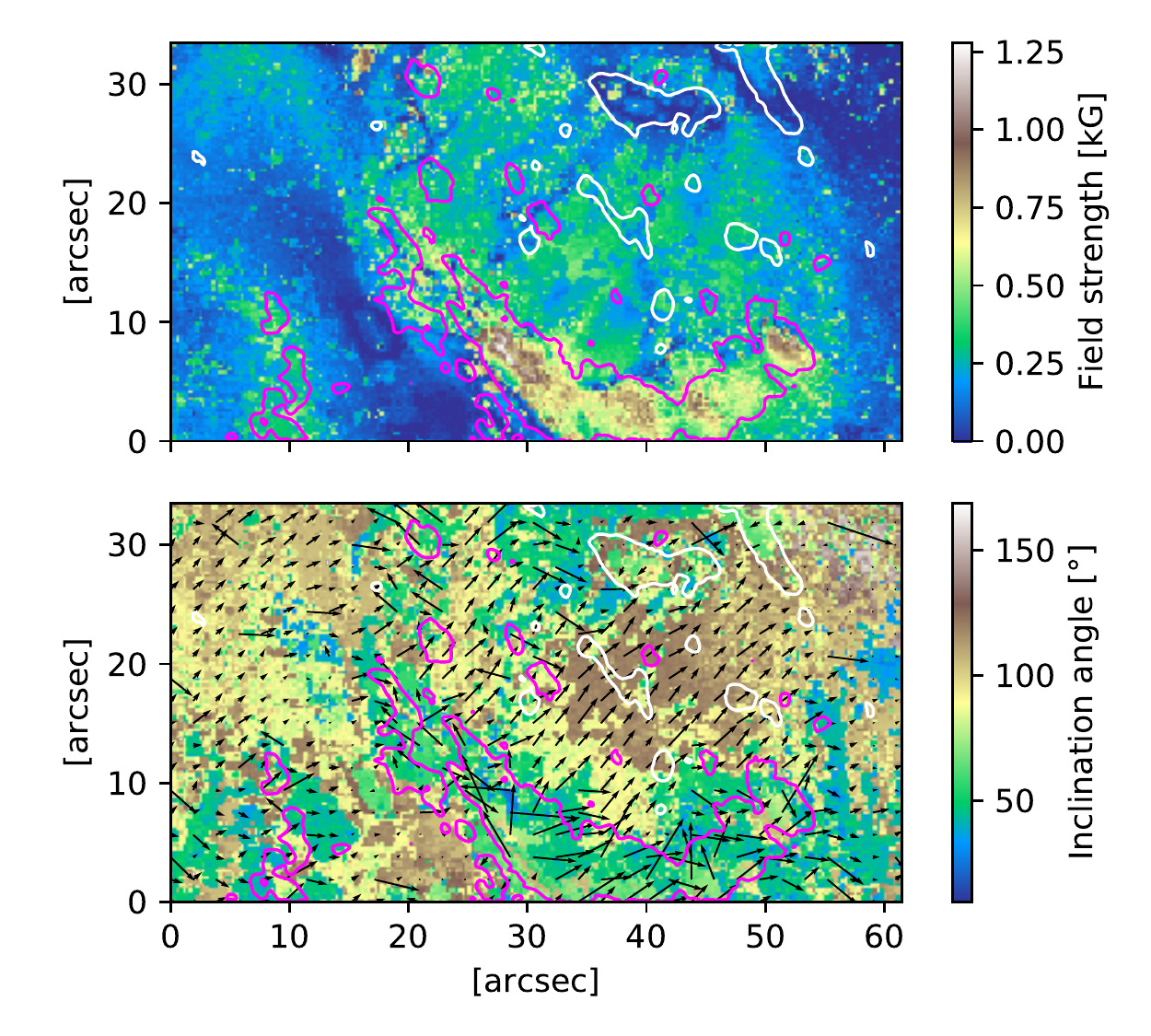}
\caption{Top: field strength retrieved by inverting He~\scriptsize{I}\normalsize{}~triplet line. Bottom: inclination angle (in background) and horizontal magnetic field vectors are shown as black arrows. The closed white and magneta contours represent negative and positive polarities in FOV at the photospheric level.}
\label{chromo_fld_inc}
\end{figure}

\subsection{Ambiguity resolution}
As mentioned above, at the photospheric height the transverse component of magnetic field vector is affected by well-known 180$^{\circ}$ azimuth ambiguity, which can be resolved using different available methods. In addition to this, the chromospheric azimuthal map inferred from Hazel can have multiple possible solutions, induced by both the Zeeman and the Hanle effects. These solutions depend on the specific regime of the magnetic field and on the scattering geometry of the FOV \citep{Schad2015, 2019A&A...625A.128D}. The number of solutions can go up to four (or up to eight if the Stokes $V$ signal is under the noise or not given), which can be resolved using different approaches provided by various authors \citep{2003Natur.425..692S,2014A&A...566A..46O,Schad2015,Gonz_lez_2015}.

\begin{figure}[!t]
\centering
\includegraphics[width=0.49\textwidth]{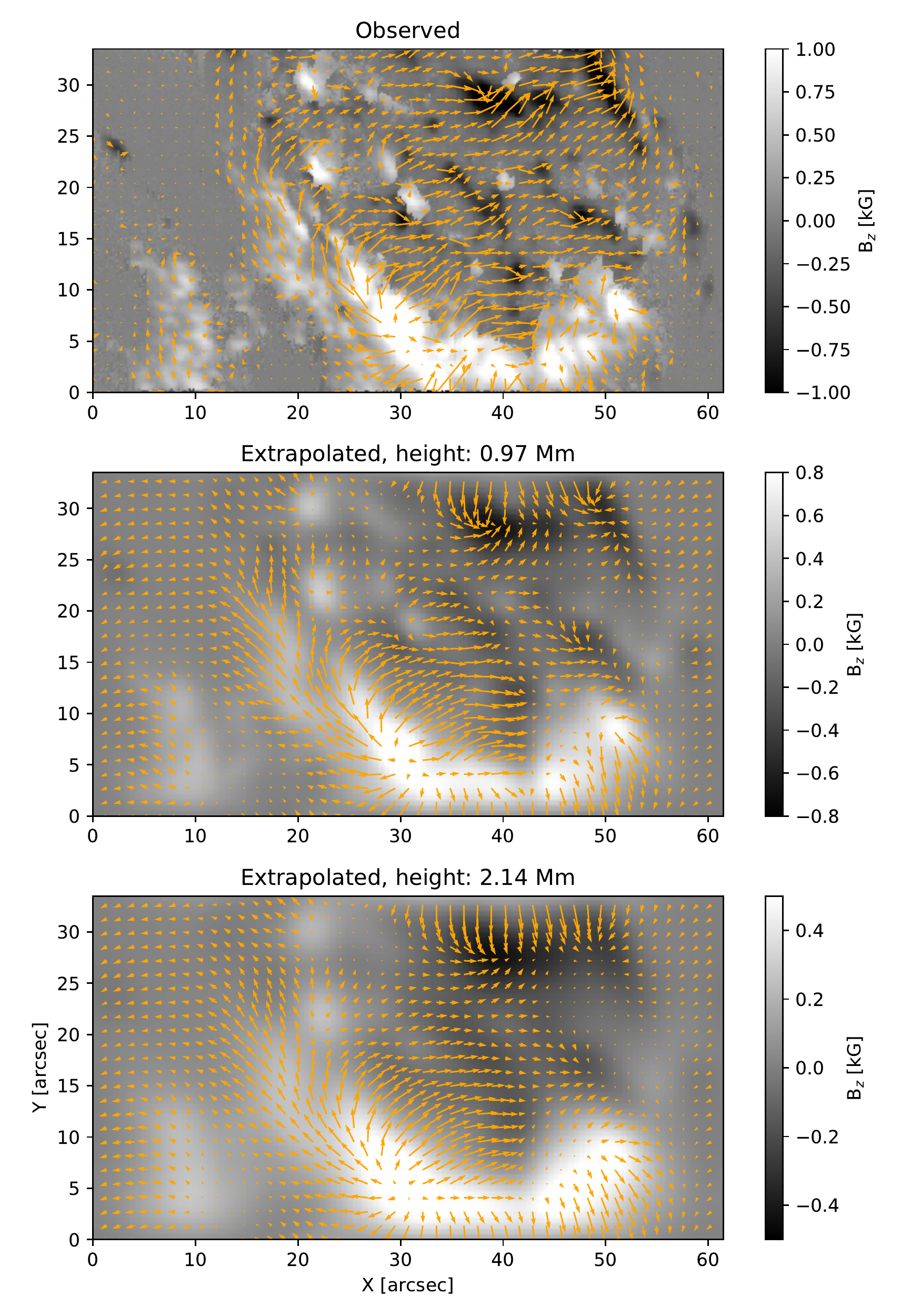}
\centering
\caption{Top panel: The observed B$_z$ components of the photospheric magnetic field derived from the Milne-Eddington inversion of Si I line. Middle and bottom panels: B$_z$ components of the photospheric magnetic field constructed using the NFFF extrapolation at different heights. The arrows show the orientation of magnetic field vector, where their length is proportional to the horizontal field strength.}
\label{nfff}
\end{figure}

To resolve the ambiguity in the \ion{He}{i} triplet line we have used the azimuth angle of the photospheric extrapolations at a particular height as an initial estimation for our inversion code. 
This height is estimated by comparing the photospheric extrapolated magnetic field strength (at different heights) with the inferred chromospheric magnetic field. After comparison, we find that the two maps show maximum similarity around $\sim$2~Mm from the solar surface. Then, using only the direction (azimuth angle, say $\phi_{ex}$) of the extrapolated field lines (at $\sim$2~Mm) in initial guess model, we again invert the Stokes profiles. During inversion, the azimuth angles are allowed to vary in a range ($\phi_{ex}\pm$20$^{\circ}$) so that the solution lie close to $\phi_{ex}$. Then, a visual inspection was also performed between the photospheric and chromospheric magnetograms, which reveals an overall good agreement (see Fig.~\ref{chromo_fld_inc} \& \ref{nfff}). The obtained similarity suggests that the ambiguity resolution for He line is reliable. Note that the chromospheric maps shown in Fig.~\ref{photo_chromo} and \ref{chromo_fld_inc} are obtained using a single slab. Subsequently, the ambiguity corrected inverted maps, both for the photosphere and the chromosphere, are analyzed in this study. 

\begin{figure}[!ht]
\centering
\includegraphics[width=0.49\textwidth]{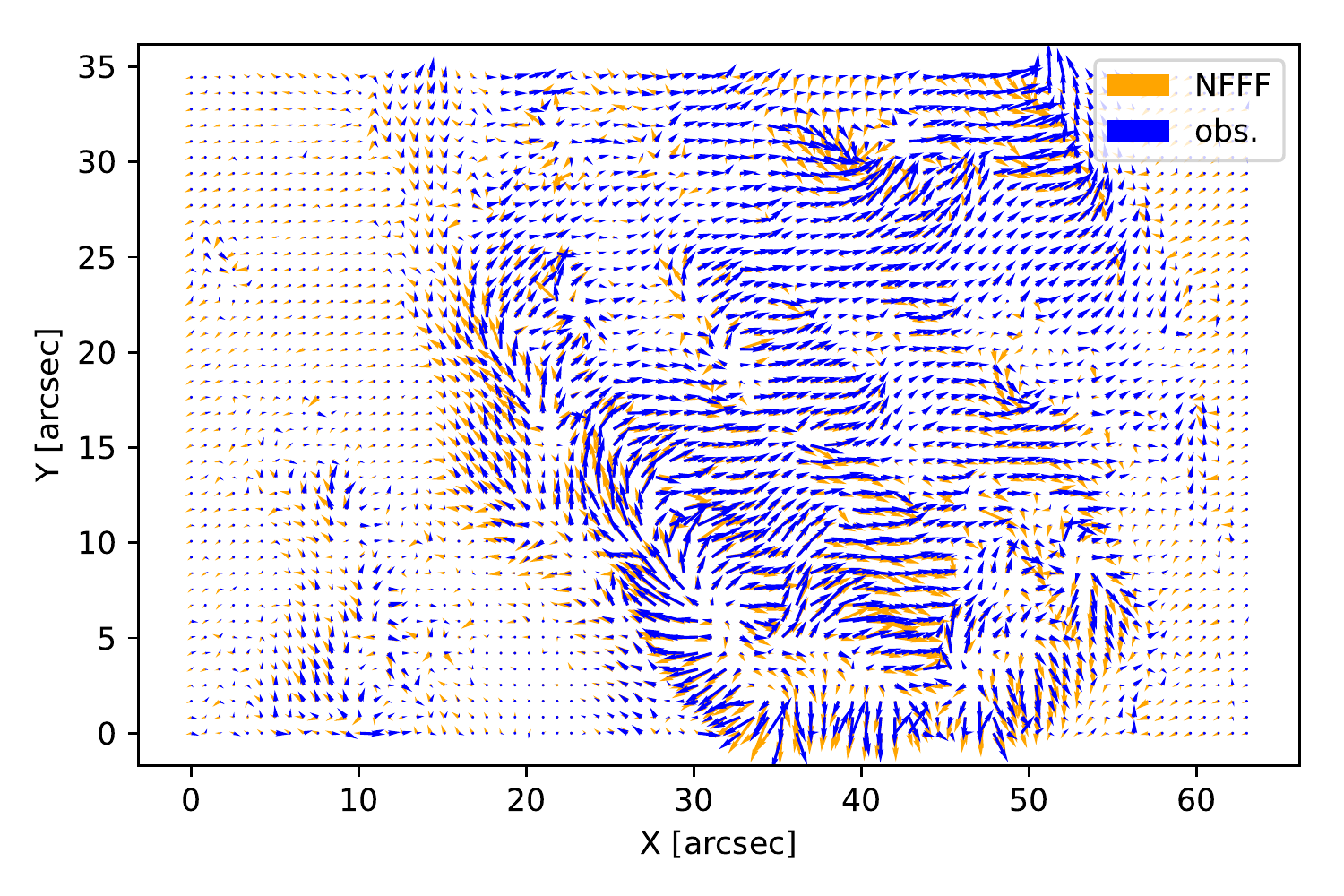}
\centering
\caption{Comparison between the observed and extrapolated horizontal magnetic field. Blue arrows represent the observed horizontal magnetic field derived using Si~\scriptsize{I}\normalsize{}~line, whereas orange arrows represent the horizontal magnetic field obtained from extrapolations at the photospheric boundary (z = 0).}
\label{obs_nfff}
\end{figure}

\section{Non-force free extrapolations}
\label{sec_nfff}

The plasma $\beta$ parameter, which is the ratio of thermal to magnetic pressure, can be of the order of unity in the photosphere \citep{Gary2001}. This indicates that the plasma in the photosphere can be in a non-force free state. Thus, in the photosphere, the retrieved current density and magnetic energy in force-free extrapolations would not be reliable \citep{2015A&A...584A..68P}. 
In order to understand the magnetic field topology more correctly in FER, we modelled the magnetic field using non-force free field extrapolations (NFFF; \citealt{ 2008SoPh..247...87H, 2010JASTP..72..219H}). The NFFF code is based on the principle of minimum dissipation rate (MDR), which is derived from a variational problem \citep{doi:10.1063/1.2828539}. Although the full MDR-based approach requires successive two
layers of vector magnetograph measurements on the solar surface, the same code can be modified for the single layer as demonstrated by \cite{2010JASTP..72..219H}. This code constructs the magnetic field lines as the superposition of two linear force free fields (LFFFs) and one potential field, $\mathbf{B}~=~\mathbf{B}_1+~\mathbf{B}_2+~\mathbf{B}_3$, where each $\mathbf{B}_i ~(i=1, 2, 3)$ obeys the following equations, 
 \begin{equation}
  \label{eq_forcefree}
      \nabla \times \mathbf{B} =  \alpha \mathbf{B};\enspace \nabla \cdot \mathbf{B} = 0,
  \end{equation}
here $\alpha$, the so-called force-free parameter, is a function of space. Without a loss of generality, we can set one of the three sub-fields  (say, $\mathbf{B}_1$) to be potential by choosing $\alpha_1$ =0. While other non-potential pair satisfies, $\alpha_{2} \ne \alpha_{3}$ $\ne$ 0. Then, the optimal pair ($\alpha_{2}, \alpha_{3}$) is obtained by minimizing the following quantity,

\begin{equation}
    E_n =\frac{\sum_{i=1}^{M}|\mathbf{B}_{t,i} - \mathbf{b}_{t,i}|}{ \sum_{i=1}^{M}|\mathbf{B}_{t,i}|},
\end{equation}
where M represents the total number of grid points on the transverse plane. Further, $\mathbf{b}_t$ and $\mathbf{B}_t$ are the computed and measured transverse magnetic field vectors on the bottom boundary, respectively. Further details of the code and its application to the solar active regions are given in \cite{ 2008SoPh..247...87H} (see also \citealt{2010JASTP..72..219H})

\subsection{Extrapolation using photospheric $\&$ chromospheric magnetogram}
In this section we describe the extrapolations of magnetic field lines using the magnetograms observed in the photosphere and the chromosphere. We first employ the photospheric vector magnetogram, retrieved by inverting the Si~\scriptsize{I}\normalsize{}~10827 line using M-E approximation, as the lower boundary condition for NFFF extrapolations. The computation are performed in a box of 226~$\times$~125~$\times$~125 grid points in the x, y, and z directions, respectively. The distance between two grid points in the horizontal and vertical directions is 0\arcsec.27~($\sim$ 200~km). For the photospheric extrapolation the obtained $E_n$ value is 0.07, suggesting the observed and the extrapolated magnetograms at the photospheric level (z = 0) are remarkably similar. As an example of extrapolations starting from the photosphere, Fig.~\ref{nfff} (top panel; in background) shows the observed B$_z$ component of the photospheric magnetic field, whereas the middle and bottom panels depict the extrapolated B$_z$ components of the constructed magnetic field using the NFFF at different heights. In each panel, the arrows represent the direction of magnetic field vector, where their length is proportional to the horizontal field strength. This figure also demonstrates that for the higher regions the magnetic field strength decreases and the field vector becomes smoother, which is generally observed in extrapolations. The similarity between the observed and extrapolated horizontal magnetic field is shown in Fig.~\ref{obs_nfff}. In this figure, blue arrows represent the observed horizontal magnetic field derived from the inversion of \ion{Si}{i} line, whereas orange arrows represent the horizontal magnetic field field obtained from extrapolations at the photospheric boundary (z = 0). It is evident from the figure that the observed and computed extrapolations are remarkably similar at the lower boundary (z = 0), yielding a linear Pearson correlation coefficient value of 0.97. This comparison also ensure us that for the \ion{Si}{i} line the ambiguity correction is reliable.

Although various authors have used extrapolations starting from photospheric magnetogram, there are only few studies performed to understand the magnetic topology using the chromospheric magnetogram as lower boundary condition. Recently, using the photospheric and the chromospheric magnetograms as lower boundary condition, two independent NLFFF extrapolations were performed by \cite{2012ApJ...748...23Y} to understand the magnetic field topology of an active region filament. In addition to photospheric extrapolation, we also performed independent extrapolations using a chromospheric vector magnetogram, which is inferred by inverting the helium triplet line, as lower boundary condition. We use the same computational domain as described above for the photospheric extrapolations. The $E_n$ value for this extrapolation was 0.56. The higher value in this case as compared to the photospheric field is expected as the field strength is weaker in the chromosphere and the uncertainities in the transverse field are larger. The observed magnetogram is relatively small and not isolated from the active region. So the region is far from the flux-balance condition, which is one of the assumptions used in the extrapolations. Therefore, the extrapolated results near the boundary can render inappropriate topology of field lines. Thus, we mainly consider the region a few arcseconds away from the FOV boundaries. The extrapolated field lines computed using the observed chromospheric magnetic field as lower boundary condition are shown in Fig.~\ref{top-side_extrapl}.


\section{Results}
\label{sec_results}
\subsection{Magnetic field in the photosphere and chromosphere}
The retrieved magnetic parameters in the FER, using the \ion{Si}{i} line, exhibit a complex magnetic structure in the photosphere. The B$_z$ component of magnetic field in photosphere illustrates that there are a few magnetic bipolar features (MBFs) in FER, two of them are highlighted in box A $\&$ B (depicted in Fig.~\ref{photo_chromo}). The polarity of MBFs close to a pore is opposite to the polarity of pore itself. In contrast to the photospheric structure, the magnetic field in the chromosphere is relatively smooth and exhibits no signature of MBFs. The latter suggests that the MBFs features are more prominent in the photosphere rather than the chromosphere, which is consistent with results from previous studies \citep{2002SoPh..209..119B,2003Natur.425..692S,Xu2010} 
Furthermore, the magnetic field inferred in the photosphere and in the chromosphere displays a weak and horizontal magnetic field in the FER. However, in the photosphere, relatively stronger and vertical magnetic fields are located in pores ($\sim$1300~G) and in MBFs ($\sim$800~G), whereas in the vicinity of FER it varies from 350~G to 600~G. Generally, in the solar atmosphere, the magnetic field strength decreases with height. For instance, the mean field strength value of a pore ($\sim$1350~G) observed in the photosphere (located at X = 30\arcsec, Y = 7\arcsec) drops to 880~G in the chromosphere. Whereas the magnetic field value ranging from 100 to 400 G in FER decreases by a factor of $\sim$0.3--0.6 compared to the photosphere. If the average distance between the \ion{Si}{i} and \ion{He}{i} formation heights is assumed as 1000--1500 km \citep{1994A&A...287..229S,2007ApJ...671.1005B,Xu2010,2016A&A...596A...8J}, then the vertical gradient ($\Delta \mathrm{B}/\Delta \mathrm{z}$) in the FER turns out to be $\sim$0.2 G km$^{-1}$, which is consistent with the values presented by different authors \citep{Xu2010,2016A&A...596A...8J}. 


\subsection{LOS Velocity in photosphere and chromosphere}
The inferred LOS velocity in the photosphere and chromosphere are shown in Fig.~\ref{photo_chromo}. In the photosphere, between the opposite polarity pores or around the mixed-polarity regions the plasma show upflows and downflows. These flows demonstrate the expansion of the FER in the photosphere when the pores push through the surrounding plasma. Downflows (up to 3.3~km s$^{-1}$) are generally located near the pores of positive polarity, whereas upflows (up to 2.2~km s$^{-1}$) are located near the negative pore or negative polarity region. In contrast to the photosphere, LOS velocities in the upper chromosphere are relatively smooth. Near the foot-points of the elongated dark loops, connecting pores of opposite polarities, we found strong downflows (reaching supersonic velocities of 40 km/s), whereas strong upflows (of 22~km/s) are observed near the middle part of the loop.

As an example, the intensity profiles of the He triplet line along a dark featured loop are shown in Fig.~\ref{stokesi_profiles}, where the values in parentheses denote the selected position of the profiles. The top profile represent the loop apex, whereas the bottom profile represent the loop foot-point. The loop apex is strongly blueshifted (upflows). Moreover, this blueshift gradually shifts towards redshift (downflows) as a function of decreasing loop height. Near the footpoints the profiles are strongly redshifted showing two velocity components. The presence of two components of velocity near the location of loop foot-points are also observed by \cite{2007A&A...462.1147L, Xu2010, 2018A&A...617A..55G}. The strong upflows near the tops of freshly emerged loops suggests that they carry cool material to the upper atmosphere, which is not yet heated to high temperature. Also strong downflows near the foot-points suggest that the loop is still rising and chromospheric cool gas is draining through the loop legs, which is consistent with the interpretation from previous studies \citep{2000ApJ...544..567S, 2003Natur.425..692S, Xu2010,2018A&A...617A..55G}. 

The overall chromospheric velocity map is remarkably different from the photospheric velocity map. However, in contrast to \cite{2003Natur.425..692S, 2007A&A...462.1147L,2018A&A...617A..55G}, we observed photospheric downflows near the location of strong chromospheric downflows. The maximum photospheric downflows observed at these location is 3.3\,km\,s$^{-1}$, which is around twice the value reported by \cite{Xu2010}.

\begin{figure*}[!t]
\centering
\includegraphics[width=1\textwidth]{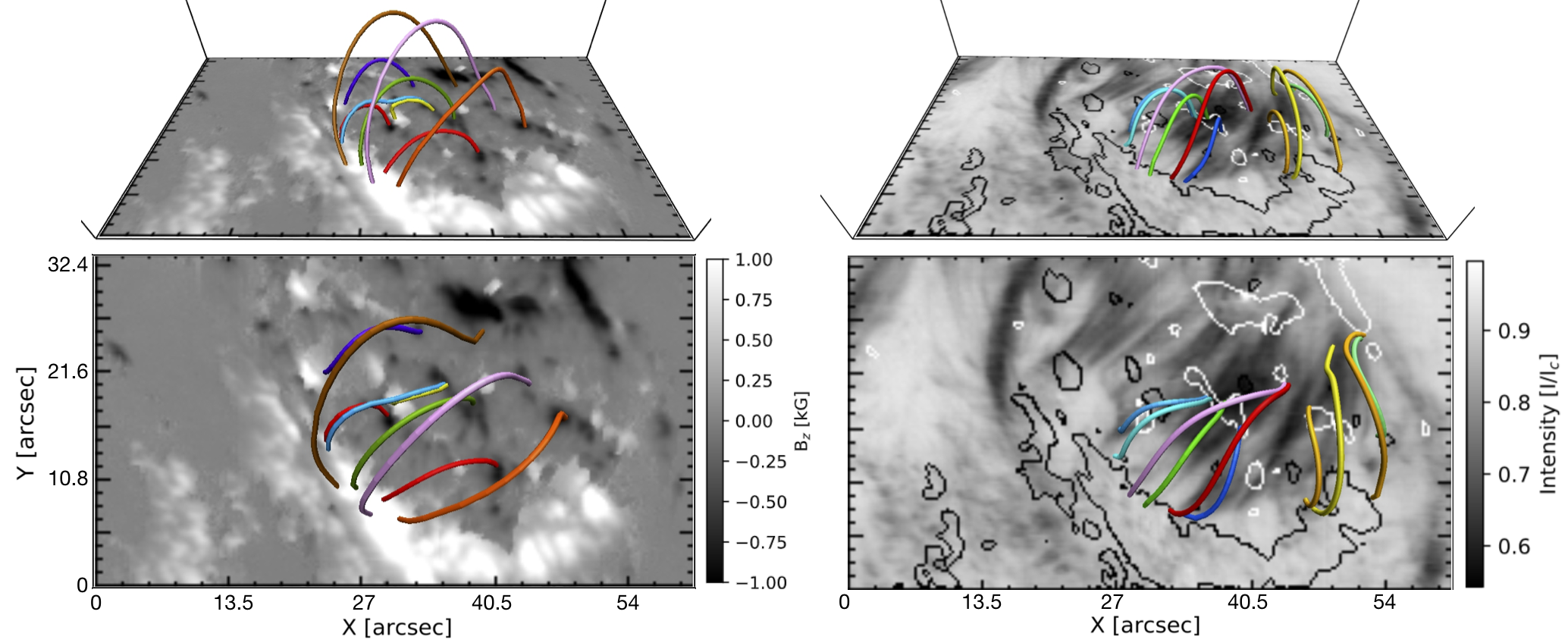}
\caption{Three dimensional view of magnetic field lines. Left panels: the extrapolated field lines computed using the observed photospheric magnetic field as lower boundary condition (top panel: lateral view, bottom panel: top view). In background B$_z$ component of magnetic field strength is shown.  Right panels: the extrapolated field lines computed using the observed chromospheric magnetic field as lower boundary condition (top panel: lateral view, bottom panel: top view). In background the intensity image shows He absorption features along the loops, closed black and white contours represent the locations of photospheric positive and negative polarities, respectively.}
\label{top-side_extrapl}
\end{figure*}

\begin{figure}[!th]
\centering
\includegraphics[width=0.45\textwidth]{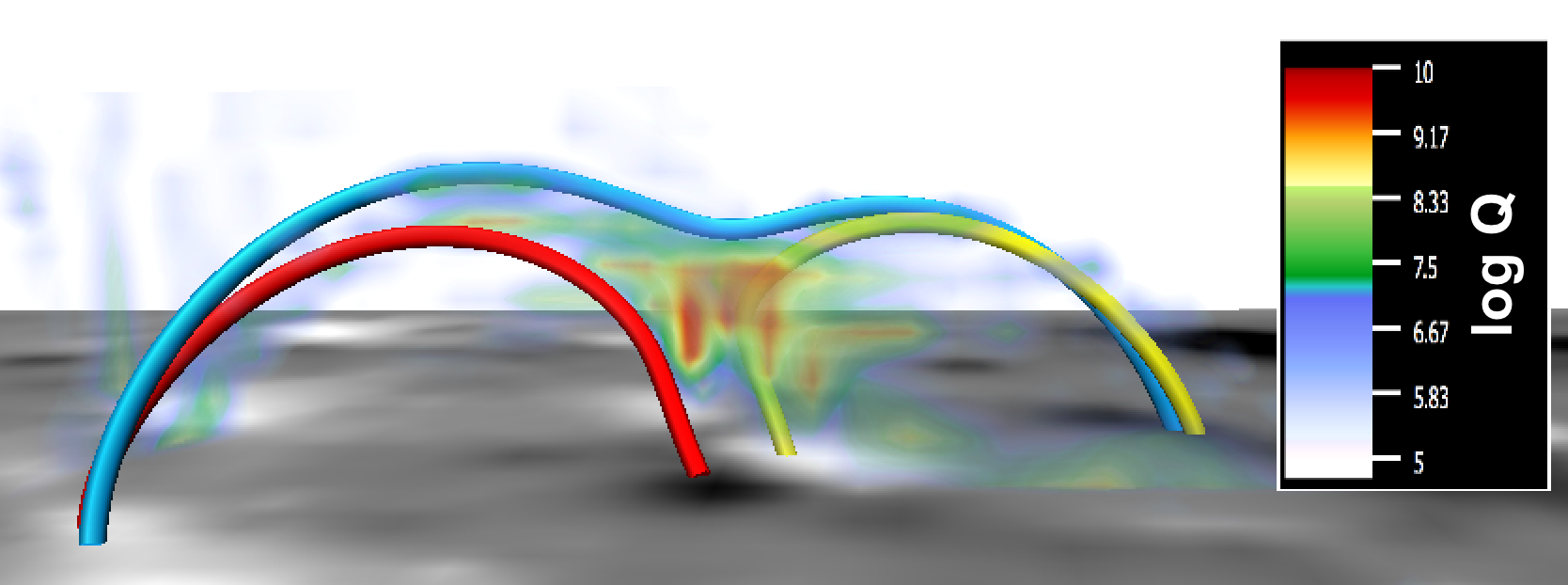}
\includegraphics[width=0.45\textwidth]{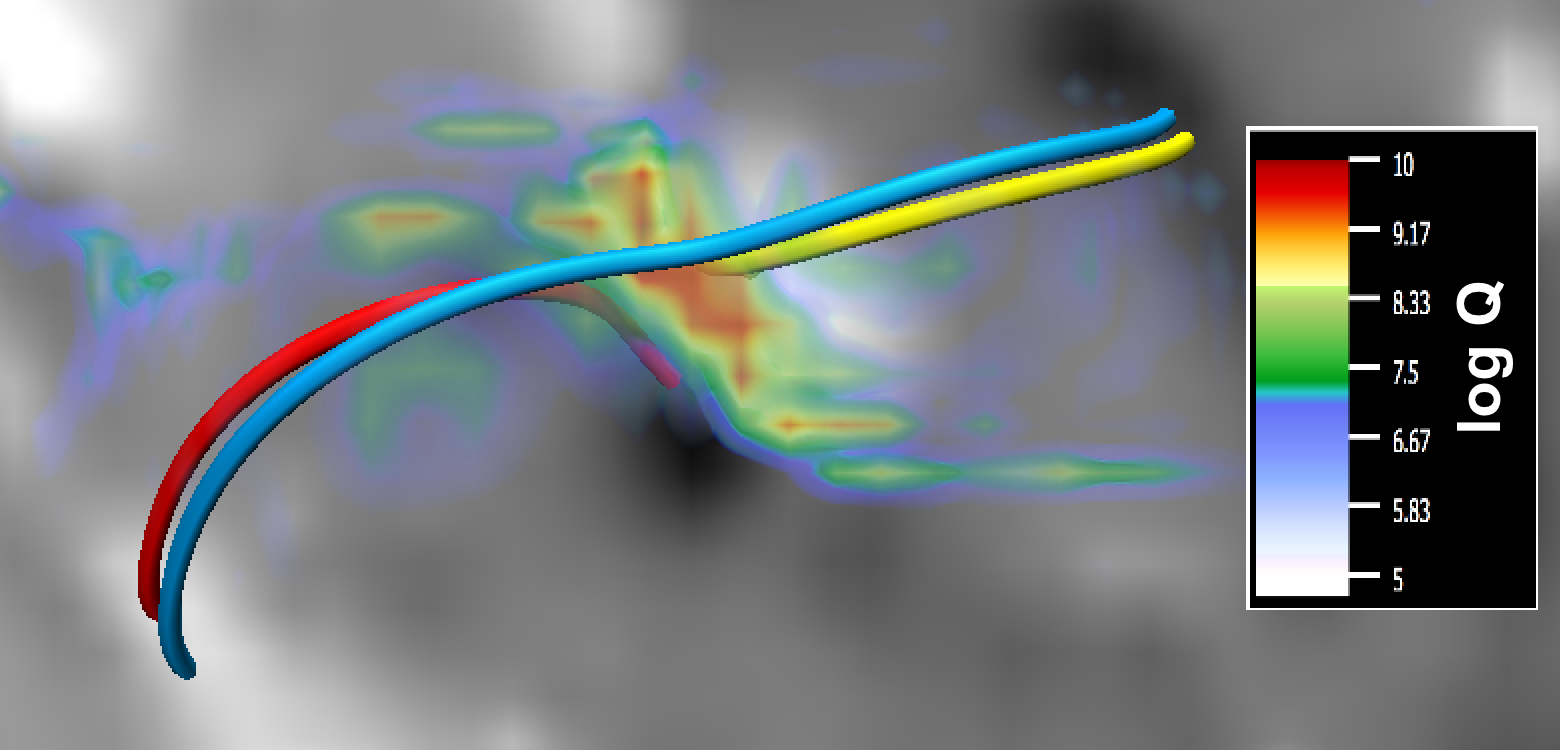}

\caption{Top (oblique view) and bottom (top view) panels display the squashing factor around a mixed polarity regions indicated by box B in Fig. \ref{photo_chromo}. In background black and white represent the locations of positive and negative polarity in the photosphere. The closed field lines connecting opposite polarities are shown in different colors.}
\label{top-side_logQ}
\end{figure}

\begin{figure}[th]
\centering
\includegraphics[width=0.49\textwidth]{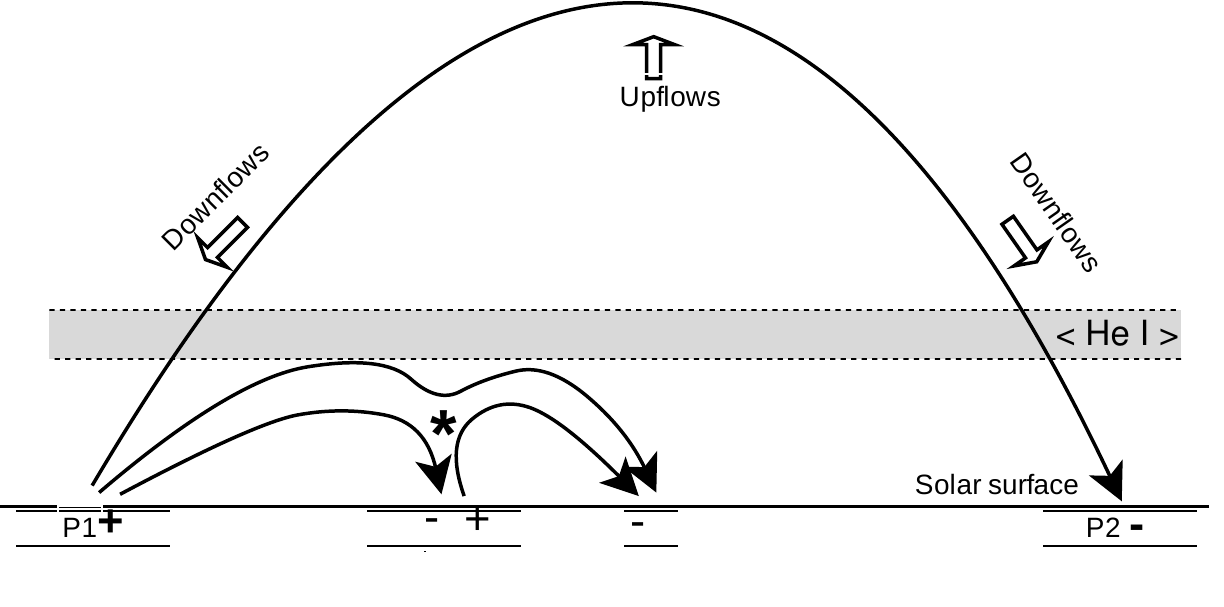}
\caption{A sketch of the magnetic field configuration in a FER based on the two layer extrapolations. The '+' and '-' symbols represent positive and negative polarity on solar surface, respectively. The gray area outlined by dashed lines represent the average formation height of \ion{He}{i} layer. An AFS connecting two pores (P1 and P2) of opposite polarity is represented by a black closed loop. Arrows indicate the upflows and downflows of plasma along an AFS as observed in the He triplet line. The asterisk symbol indicates the plausible location of magnetic reconnection near a mixed polarity region.}
\label{afs_cartoon}
\end{figure}

\subsection{Magnetic field topology and extrapolated field lines}
In addition to the reconstruction technique described by \cite{2003Natur.425..692S}, the field lines can also be reconstructed using extrapolation techniques as demonstrated in \cite{2005A&A...433..701W}. Therefore, in order to understand the magnetic field topology in FER, we also reconstructed the magnetic field lines using extrapolation approximations. The three dimensional view of photospheric/chromospheric extrapolated field lines in the FER are presented in Fig.~\ref{top-side_extrapl}. Although open and closed field lines are present in the FOV, for clarity we depict only selected closed field lines.
The figure demonstrates that a \textit{normal} magnetic field configuration is obtained, where the field lines start from a positive polarity and stop near a negative polarity regions. From a visual inspection we note that almost all closed extrapolated field lines are parallel to the elongated dark features (see also \citealt{2015ApJ...802..136L} for the relation between H$\alpha$ fibrils and the magnetic field lines in a three-dimensional MHD simulation of a network region). 

Using the photospheric extrapolations we have estimated the average formation height of the He line in the FER. We first compared the magnetic field strength in the photospheric extrapolation cube at various heights with the inferred chromospheric one. We found that they show maximum correlation around $\sim$1.5~Mm. If we assume that the averaged formation height of the \ion{Si}{i} line in a FER is 0.5~Mm from the solar surface ($\tau$ = 1; \citealt{2008ApJ...682.1376B}), then, the averaged formation height of the \ion{He}{i} turns out to be 2~Mm (1.5~Mm + the formation height of the \ion{Si}{i} line) from the solar surface.

Noticeably, many field lines, reconstructed using photospheric and chromospheric extrapolations, start and end around same locations. Under normal field configurations the foot-points of field lines seen in the photosphere and the chromosphere should lie around the same polarity regions, which is supported by our extrapolations. The maximum height of loops reconstructed using photospheric extrapolation is 10.5~Mm above the solar surface with the foot-point separation of 16~Mm. Furthermore, as shown in Fig.~\ref{top-side_extrapl}, there are several small loops connecting pores to nearby magnetic structures of opposite polarity. For instance, one of the small loops have a footpoint separation of 5~Mm and is $\sim$2~Mm high. 

On the other hand, the magnetic loops reconstructed using the chromospheric extrapolations exhibit similar topology of field lines. They show that the maximum height of a reconstructed loop along an AFS is around 8.4~Mm (or 6.4~Mm from the chromospheric height) with the foot-point separation of 16~Mm at the chromospheric height. In our case the extrapolated loops are comparable to the magnetic loops reconstructed by \cite{2003Natur.425..692S}. They reconstructed loops with a foot-point separation of 20~Mm and 10~Mm high. In addition to this, using the same approach demonstrated by \cite{2003Natur.425..692S}, in a different FER, \cite{Xu2010} have also reconstructed loops of 4--4.5~Mm high from the solar surface, which is consistent with the present study. Since we analyzed only limited FOV of FER, the higher loops may exist in the full FER. Further, the loop height heavily depends on strength of $\mathbf{B_z}$, which may vary from one FER to another.

The photospheric extrapolations reveal that strong electric currents form around mixed polarity regions. They normally appear at the locations of strong magnetic field gradient. The squashing factor (Q), generally used to quantify this gradient, shows the most probable location of magnetic reconnections to occur. Therefore, we also calculate the three-dimensional Q maps using the method of \cite{2016ApJ...818..148L}. As an example, we show top and oblique view of Q map at the location of a mixed polarity region (box B in Fig. \ref{photo_chromo}) in Fig. \ref{top-side_logQ}, which are generated using the photospheric extrapolations. We observe that intense brightening in AIA/SDO 1600 and 1700 \AA~maps are associated with large Q values, which could be EBs or UV burst \citep{2004ApJ...614.1099P,2015ApJ...812...11V,2018SSRv..214..120Y}. Unfortunately, to confirm these features as EBs or UV burst we did not have much information from other spectral lines. 
It is evident from the Fig. \ref{top-side_logQ} that large Q values are located on the photospheric inversion line, also known as bald patches \citep{1993A&A...276..564T}, and they lie in region of 1 -- 1.6~Mm high from the solar surface. On the other hand, using chromospheric extrapolations we did not observe strong currents and signatures of magnetic reconnection in the chromosphere. The analysis of the magnetic topology and Q maps suggests that the observed brightening are generated through magnetic reconnection in the lower chromosphere or upper photosphere, which is consistent with the previous suggestions \citep{2002ApJ...575..506G,2004ApJ...614.1099P,2013ApJ...774...32V,2015ApJ...810..145D,2017ApJ...836...63T,2018ApJ...854..174T,2019A&A...627A.101V}.

\section{Discussion and conclusions}
\label{sec_conclusion}
In this article we have analyzed spectropolarimetric observations of a young FER located close to the solar limb. We have analyzed the magnetic and kinematic nature of a FER in the photosphere and chromosphere using the \ion{Si}{i} 10827~\AA~and \ion{He}{i} 10830~\AA~triplet lines. In order to retrieve the physical properties of solar atmosphere, the \ion{Si}{i} line is inverted using Milne-Eddington atmosphere, whereas \ion{He}{i} triplet line is inverted using the Hazel inversion code by considering a joint action of the Hanle effect and the Zeeman effect. 

Through spectropolarimetric analysis of the \ion{Si}{i} line we observed a complex magnetic structure showing  a few  magnetic bipolar elements near the vicinity of FER. In the photosphere, the magnetic field is weak (350 -- 600 G) and horizontal around the middle part of FER. However, it becomes relatively strong ($\sim$1000 G) and more vertical near  the pores. The obtained overall features at the photospheric level are consistent with previous results \citep{2003Natur.425..692S, 2004A&A...414.1109L, Xu2010}. Our analysis of the \ion{He}{i} triplet line shows a smooth variation of the magnetic field vector (ranging from 100~G to 400~G) and velocity across FER. Furthermore, we find supersonic downflows of $\sim$40~km~s$^{-1}$ near the foot points of loops connecting two pores of opposite polarity, while a strong upflows of $\sim$22~km~s$^{-1}$ near the middle part of loops. Recently, \cite{2018A&A...617A..55G} studied the evolution of an AFS in a FER using 64 minutes of observations. Using \ion{He}{i} 10830 intensity profiles, they measured strong upflows (near the central part) and downflows (near the foot-points) in loops. During 30 minutes of lifetime, the upflows decreases gradually whereas the downflows increases in the middle stage and again decreases at the later stage. Finally, the AFS vanishes when downflows drops to zero. Based on their observations, the strong upflows in the present study indicates that the loops are young and carry cool material to the upper atmosphere, which is also evident from the strong \ion{He}{i} triplet line absorption along the emerged loops. 

In contrast to \cite{2007A&A...462.1147L, 2018A&A...617A..55G}, at the location of the supersonic downflows in the chromosphere, we observed downflows of $\sim$3 km s$^{-1}$ in the photosphere. In addition to this, \cite{Xu2010} have also reported photospheric downflows of 1.5 km s$^{-1}$ below regions with chromospheric downflows. The observed strong photospheric downflows could be a consequence of atmospheric disturbances caused by the supersonic downflowing material along the loop foot-points. Moreover, in contrast to \cite{2007A&A...462.1147L}, no signs of emission in the \ion{He}{i} line observed in our analysis, which indicates that the transition of downflowing material from supersonic to sonic velocities lie below the \ion{He}{i} line formation. 

To understand the magnetic field topology in FER, we employed NFFF extrapolations starting from our photospheric and chromospheric magnetic field measurements. A \textit{normal} magnetic configuration (connecting positive to negative polarity regions) is observed above the FER. We observed that there are several magnetic loops of small and large scales, where small-scale loops lie beneath the large-scale loops forming magnetic canopies of various size. The magnetic loops reconstructed using photospheric extrapolations along an AFS have a maximum height of 10.5~Mm with the foot-points separation of 19~Mm, whereas the loops reconstructed using chromospheric extrapolations are around 8.4 Mm high with the foot-point separation of 16~Mm.
They are almost aligned along the \ion{He}{i} line absorption features, which is consistent with the previous findings. However, in the lower chromosphere, the observations in \ion{Ca}{ii}~8542~\AA~have revealed that the magnetic vectors are mostly directed along the dark features or fibrils, but not always \citep{2011A&A...527L...8D, 2017A&A...599A.133A}, which could be an effect of the \ion{Ca}{ii}~8542~\AA~line forming deeper in the atmosphere. 

The small-scale heating events (EBs or UV bursts) and energy release at different heights have been observed above the FER. Our analysis suggests that the appearance of these events, or in general the heating of upper layers in the solar atmosphere, is likely due to the magnetic reconnection between the magnetic loops of opposite directions. We also observed that the intense small-scale bright features in AIA 1600 and 1700 \AA~are associated with large Q values, indicating that they are energized by small scale magnetic reconnections, which is consistent with the previous suggestions.
In our analysis, the magnetic field strength is horizontal in the vicinity of FER, but no significant brightening observed in the upper chromosphere using \ion{He}{i} triplet line. However, in a recent analysis of FER using multi-line observations at high spatial and temporal resolution, \cite{2018A&A...612A..28L} reported that the heating rate in the low chromosphere correlates with the strength of the horizontal magnetic field. The role of magnetic field vector in the chromospheric heating is still unclear, thus more investigations of FER or active region using multi-line observations are required.


Based on the analysis of spectropolarimetric data at two layers, the magnetic field configuration of a FER is sketched in Fig.~\ref{afs_cartoon}. This cartoon also illustrates the rearrangement of magnetic field lines during reconnection near a mixed polarity region, which lie in the lower chromosphere or upper photosphere. The overall magnetic topology in the FER shown here is in agreement with the previous findings \citep{2003Natur.425..692S,Xu2010, 2017ApJ...836...63T,2018A&A...617A..55G} and supports the theoretical AFS models. In future we would investigate the magnetic configuration of a FER and the presence of current sheets around the loops using two boundary conditions (at different layers) in a single extrapolation. Moreover, multi-line observations at high-spatial and temporal resolution required to understand the observational consequences of FER in different layers. The advent of new generation solar telescopes, such as, the Daniel K. Inouye Solar Telescope (DKIST; \citealt{2015csss...18..933T}) and the European Solar Telescope (EST; \citealt{2016SPIE.9908E..09M}) would be crucial in this regard.

\begin{acknowledgements}
We would like to thank the anonymous referee for the comments and suggestions. The 1.5 m GREGOR solar telescope was built by a German consortium under the leadership of the Kiepenheuer Institut für Sonnenphysik in Freiburg with the Leibniz Institut für Astrophysik Potsdam, the Institut für Astrophysik Göttingen, and the Max-Planck Institut für Sonnensystemforschung in Göttingen as partners, and with contributions by the Instituto de Astrofísica de Canarias and the Astronomical Institute of the Academy of Sciences of the Czech Republic. The observing time at GREGOR was provided by the Trans-National Access and Service Programme of the SOLARNET project, funded by the European Commission’s 7th Framework Programme under grant agreement No. 312495. JdlCR is supported by grants from the Swedish Research Council
(2015-03994), the Swedish National Space Board (128/15) and the Swedish Civil Contingencies Agency (MSB). This project has received funding from
the European Research Council (ERC) under the European Union's Horizon 2020 research and innovation programme (SUNMAG, grant agreement 759548).
The Institute for Solar Physics is supported by a grant for research infrastructures of national importance from the Swedish Research Council
(registration number 2017-00625). AP acknowledges partial support of NASA grant 80NSSC17K0016 and NSF award AGS-1650854. AAR acknowledges financial
support from the Spanish Ministerio de Ciencia, Innovaci\'on y Universidades through project
PGC2018-102108-B-I00 and FEDER funds. We acknowledge the use of the visualization software VAPOR (www.vapor.ucar.edu) for generating relevant graphics. Data and images are courtesy of NASA/SDO and the HMI and AIA science teams. This research has made use of NASA’s Astrophysics Data System.
We acknowledge the community effort devoted to the development of the following open-source packages that were used in this work: numpy (\url{numpy.org}), matplotlib (\url{matplotlib.org}) and sunpy (\url{sunpy.org}).
\end{acknowledgements}

\bibliographystyle{aa}
\bibliography{new-ref}  

\end{document}